\documentclass[journal=jacsat,manuscript=article]{achemso}
\usepackage{booktabs}
\usepackage{siunitx}
\usepackage{graphicx} 
\usepackage{mhchem}
\usepackage{subcaption}
\usepackage{float}
\usepackage{mathtools}
\usepackage{comment}
\usepackage{gensymb} 
\usepackage{svg} 
\usepackage{tocloft}

\newcommand{\dhd}{$\Delta H_{d}$}
\newcommand{\dhf}{$\Delta H_{f}$}

\newcommand{\nthree}{$A^{3+}$}
\newcommand{\nfour}{$A^{4+}$}
\newcommand{\nfive}{$A^{5+}$}
\newcommand{\nsix}{$A^{6+}$}

\author{Zachary J. L. Bare}
\affiliation{Department of Chemistry and Biochemistry, University of South Carolina, South Carolina 
29208, United States}

\author{CJ Sturgill}
\affiliation{Department of Chemistry and Biochemistry, University of South Carolina, South Carolina 
29208, United States}

\author{Manish Kumar}
\affiliation{Department of Chemistry and Biochemistry, University of South Carolina, South Carolina 
29208, United States}

\author{Iva Milisavljevic}
\affiliation{Inamori School of Engineering, Alfred University, New York 14802, United States}

\author{Hans-Conrad zur Loye}
\affiliation{Department of Chemistry and Biochemistry, University of South Carolina, South Carolina 
29208, United States}

\author{Scott Misture}
\affiliation{Inamori School of Engineering, Alfred University, New York 14802, United States}

\author{Morgan Stefik}
\email{Stefik@mailbox.sc.edu}
\affiliation{Department of Chemistry and Biochemistry, University of South Carolina, South Carolina 
29208, United States}

\author{Christopher Sutton}
\email{cs113@mailbox.sc.edu}
\affiliation{Department of Chemistry and Biochemistry, University of South Carolina, South Carolina 29208, United States}

\title{Computational screening and experimental validation of promising Wadsley-Roth Niobates}

\begin{document}

\begin{abstract}
The growing demand for efficient, high-capacity energy storage systems has driven extensive research into advanced materials for lithium-ion batteries. Among the various candidates, Wadsley–Roth (WR) niobates have emerged as a promising class of materials for fast Li$^+$ storage due to rapid ion diffusion within their \ce{ReO3}-like blocks ($n \times m \times \infty$) in combination with good electronic conductivity along the shear planes. Despite the remarkable features of WR phases, there are presently less than 30 known structures which limits identification of structure-property relationships for improved performance as well as the identification of phases with more earth-abundant elements. 
In this work, we have dramatically expanded the set of potentially (meta)-stable compositions (with \dhd{} $<$ 22 meV/atom) to 1301 (out of 3283) through high-throughput screening with density functional theory (DFT). 
This large space of compound was generated through single- and double-site substitution into 10 known WR-niobate prototypes using 48 elements across the periodic table.
To confirm the structure predictions, we successfully synthesized and validated with X-ray diffraction a new material, \ce{MoWNb24O66}. The measured lithium diffusivity in \ce{MoWNb24O66} has a peak value of 1.0$\times$10$^{-16}$ m$^2$/s at 1.45V vs Li/Li$^+$ and achieved 225 $\pm$ 1 mAh/g at 5C. Thus a computationally predicted phase was realized experimentally with performance exceeding \ce{Nb16W5O55}, a recent WR benchmark. Overall, the computational dataset of potentially stable novel compounds and with one realized that has competitive performance provide a valuable guide for experimentalists in discovering new durable battery materials.

\end{abstract}

\section{Introduction}
The growing demand for efficient, high-capacity energy storage systems has driven extensive research into advanced materials for lithium-ion batteries (LIBs). Wadsley–Roth (WR) niobates have emerged as a promising class of materials for fast Li$^+$ storage due to their unique open-tunnel framework that can accommodate a high density of Li$^+$ ions and offer facile Li$^+$ diffusion paths.\cite{yang2021wadsley,liu2023recent} Their typical lithiation via the formation of a solid solution (second-order phase transition) enables continuous expansion/contraction, which limits cracking and enables longevity. Moreover, the two-electron transfer per Nb atom (Nb$^{5+}$/Nb$^{4+}$ and Nb$^{4+}$/Nb$^{3+}$ redox couples) contributes to a high theoretical capacity and an appropriate anode voltage window in the range of 1.0-2.0 V vs Li$^+$/Li that minimizes electrolyte degradation and the formation of a solid-electrolyte interface (SEI), preventing safety issues, such as lithium dendrite formation.\cite{yang2021wadsley}

WR phases are crystallographic shear structures composed of \ce{ReO3}-type blocks of size $n\times m$, where $n$ and $m$ represent the number of corner-sharing metal oxide octahedra (\ce{MO6}) along the length and width of the blocks, respectively.\cite{krumeich2022complex,koccer2019cation} Adjacent \ce{ReO3}‐type blocks are connected infinitely ($n \times m \times \infty$) through edge‐sharing or a combination of edge-sharing and tetrahedrally coordinated metal atoms at the block corners.\cite{cava1983lithium} The sizes of the blocks are determined by the stoichiometric ratio between metals and oxygen, where more oxygen per metal results in more connections between the octahedra and larger blocks up to size 5$\times$5.\cite{koccer2019cation} Many Nb-based oxides with crystallographic shear structures have been investigated as promising high-rate anode materials, such as H-\ce{Nb2O5} \cite{song2020ultrafast}, \ce{Nb12O29} \cite{li2019nanosheet}, \ce{TiNb2O7} \cite{han20113,griffith2020titanium}, \ce{Ti2Nb10O29} \cite{deng2020synergy}, \ce{TiNb24O62} \cite{griffith2017structural}, \ce{PNb9O25} \cite{patoux2002reversible}, \ce{FeNb11O29} \cite{zheng2019fenb11o29,spada2021fenb11o29,spada2019orthorhombic,pinus2014neutron}, \ce{Nb18W8O69} \cite{griffith2020superionic}, \ce{Nb16W5O55} \cite{griffith2018niobium}, \ce{Nb12WO33} \cite{koccer2020lithium} and \ce{Nb14W3O44} \cite{koccer2020lithium}. 

Despite the remarkable features of WR phases, there are so few known that there remains significant opportunity to discover new phases, identify structure-property relationships~\cite{salzer2023structure,muhit2024comparison}, and find compositions with more earth-abundant elements. One strategy for discovering new materials is through compositional modifications, such as ion substitution or solid solutions/alloying. However, relative to the space of potential compositions, only a small number of compositional modifications have been explored, including \ce{CoNb11O29},\cite{liu2024wadsley} \ce{AlNb11O29},\cite{lou2019new} \ce{CrNb11O29},\cite{fu2018highly} \ce{GaNb11O29},\cite{lou2017ganb11o29} \ce{VNb9O25},\cite{liang2021micro} \ce{AsNb9O25},\cite{ulutagay1998niobium} \ce{HfNb24O62}, \cite{fu2020highly} \ce{Nb$_{7}$Ti$_{1.5}$Mo$_{1.5}$O$_{25}$}, \cite{green2023structural} \\ \ce{Ta$_{7}$Ti$_{1.5}$Mo$_{1.5}$O$_{25}$}, \cite{green2023structural} \ce{Ta12MoO33}, \cite{muhit2024comparison} \ce{MoNb12O33},\cite{zhu2019monb,muhit2024comparison} \ce{NaNb13O33},\cite{patterson2023rapid} \ce{Mo$_{1.5}$W$_{1.5}$Nb$_{14}$O$_{44}$}, \cite{tao2022insight} \ce{Nb14Mo3O44},\cite{myslyvchenko2021formation} \ce{TiNb6O17}, \cite{lin2015tinb} and \ce{V3Nb9O29}. \cite{wang1985lithium}
This limitation is in part due to the synthetic and characterization challenges.

Instead, we leverage computational predictions to identify stable compositions, reducing synthetic efforts on potentially viable candidates.
There are few reports of first-principles computational modeling for WR phases,\cite{wang2022fast,griffith2019ionic,koccer2019cation,koccer2020lithium,preefer2020multielectron,tao2022insight,saber2023redox,koccer2019first,saber2023chemical} partially due to the complexity of the crystal structures with large unit cells and inherent cation disorder. This complexity is further exacerbated when considering partial ion substitutions. For instance, modeling the experimentally reported composition \ce{Mo$_{1.5}$W$_{1.5}$Nb14O44} \cite{tao2022insight} requires a minimum of 2 formula units and a unit cell size of 122 atoms.

While machine-learned interatomic potentials (MLIPs) provide an efficient and typically accurate approach for calculating properties such as stability and for generating structures, a key issue arises when applying them to the systems of interest in this study—particularly the Wadsley–Roth (WR) phases are not well represented in the materials project. The lack of entries for these compounds in the Materials Project suggests that existing MLIPs may not be sufficiently accurate when applied to this chemical space. These compounds lie outside the scope of the Materials Project and have not yet been systematically explored at large compositional scales.  Moreover, trends in stability and related properties have only been examined for a limited number of compounds.

In this work, we use high-throughput DFT calculations to identify 1301 (meta)-stable new compositions (with \dhd{} $<$ 22 meV/atom) out of 3283 compositions based on single and double A-site element substitutions for 10 experimentally observed WR niobates. The key findings from this study reveal several factors that influence stability including block size and coordination preferences. Moreover, higher Nb content and maintaining similar oxidation states on the A-sites for double substitutions enhance stability in these compositions. 
These results provide valuable insights into stability trends that can guide experimental synthesis efforts toward more stable WR-niobate compounds. 

\begin{figure}[h!]
    \centering
    \includegraphics[width=10cm]{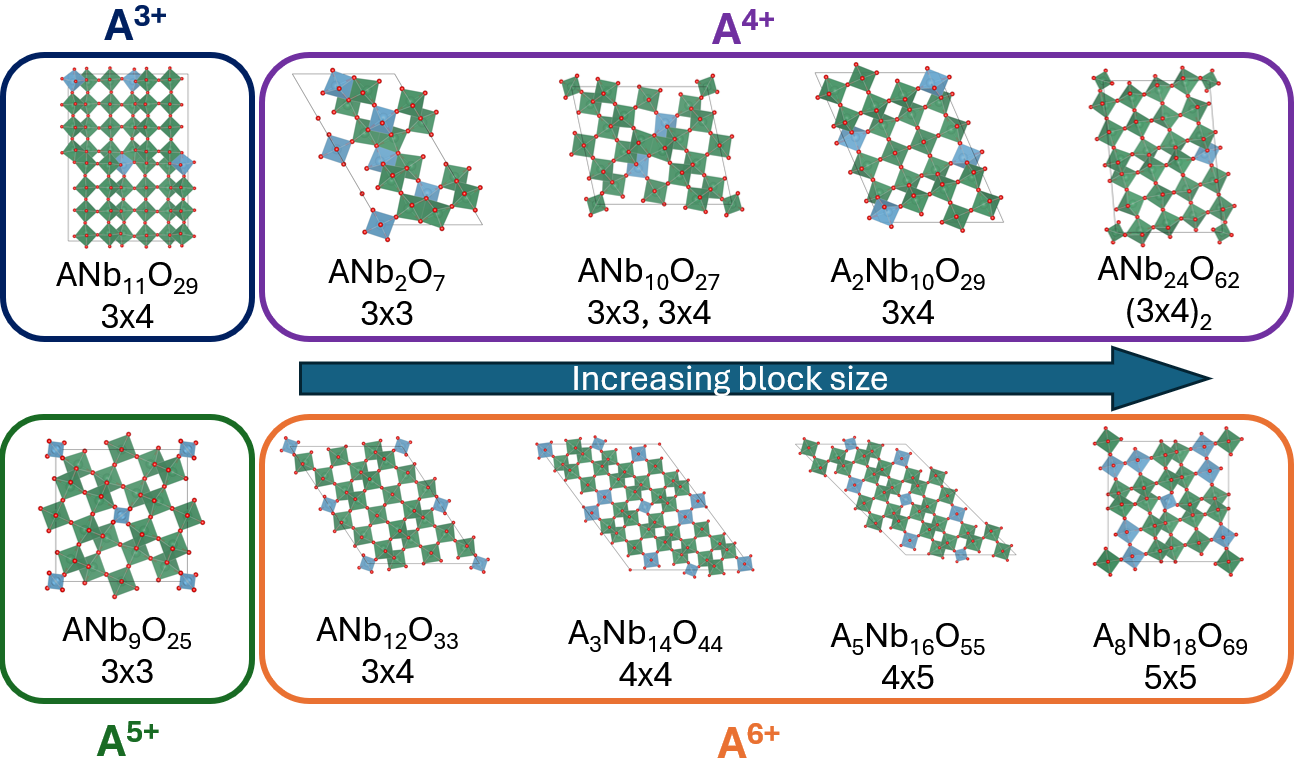}    
    \caption{Prototype structures for the 10 WR phase niobates, which were used as templates for the screening a total of 3283 compounds based on stability.}
    \label{fig:st_im}
\end{figure}

\section{Results and Discussion}

The article is organized as follows: In the first section, we discuss trends in single (isovalent) $A$-site substitutions for the 10 WR-niobate prototype structures considered in this study. The stability (decomposition enthalpy, \dhd{}) of these compounds are evaluated through a DFT-based convex hull construction (see Methods) and the trends for each $A$-site oxidation state are discussed separately for clarity. In the next section, because there are only a limited number of single $A$-site substitutions, we then discuss double $A$-site substitutions, which substantially extends the chemical space due to the number of oxidation state combinations. Within this space, we identify 1301 potentially stable compositions. In the final section, we describe the synthesis and testing of one novel compound predicted in this work and demonstrate its exceptional performance as a battery material.

\subsection{Trends in the calculated stability for isovalent $A$-site substituted compositions (single substitutions)}

This section discusses trends in single $A$-site substitutions for the 10 WR-niobate prototype structures (shown in Figure~\ref{fig:st_im}). To make it easier to follow, the various prototypes will be referred to by their prototype IDs listed in Table \ref{tab:ptype}, which are their corresponding oxygen stoichiometries based on their reduced formulae and the designated $A$-site oxidation state as a superscript (e.g., $O_{29}^{A4+}$). For each of the 10 prototype structures listed in Table \ref{tab:ptype}, 24 (\nthree), 25 (\nfour), 14 (\nfive ), and 9 (\nsix) cations (shown in SI Figure S1) were substituted into the prototype structure and subsequently relaxed with DFT, resulting in a total of 174 possible compositions. The results are organized according to the charge of the substituted cation.

To identify the substitutions that result in stable compositions, it is beneficial to compare their stabilities (\dhd{}) with that of known experimental phases. For the 22 experimentally observed phases listed in the SI Table S1, the computed \dhd{} ranges from -8 to 22 meV/atom. The least stable compositions are \ce{AlNb11O29} and \ce{W8Nb18O69} with \dhd{} of 19 and 22 meV/atom above the hull, respectively. 
Because these niobates have been previously synthesized with \dhd{} as large as 22 meV/atom above the convex hull, in this work, a compound is considered (meta)-stable (i.e. potentially synthesizable) if it has \dhd{} $< $ 22 meV/atom. Of the 174 possible compositions, 92 have \dhd{} $< $ 22 meV/atom, indicating (meta)-stability with respect to competing phases. The 92 compositions are comprised of 17 $O_{29}^{A3+}$, 10 $O_{29}^{A4+}$, 4 $O_{7}^{A4+}$, 7 $O_{27}^{A4+}$, 24 $O_{62}^{A4+}$, 9 $O_{O25}^{A5+}$, 7 $O_{O33}^{A6+}$, 7 $O_{O44}^{A6+}$, 6 $O_{O55}^{A6+}$, and 1 $O_{O69}^{A6+}$. Note that the 1 stable $O_{O69}^{A6+}$ is the experimentally observed compound \ce{W8Nb18O69} that was used to define the cutoff energy for stability.

\begin{figure}[h!]
    \centering
    \includegraphics[width=8cm]{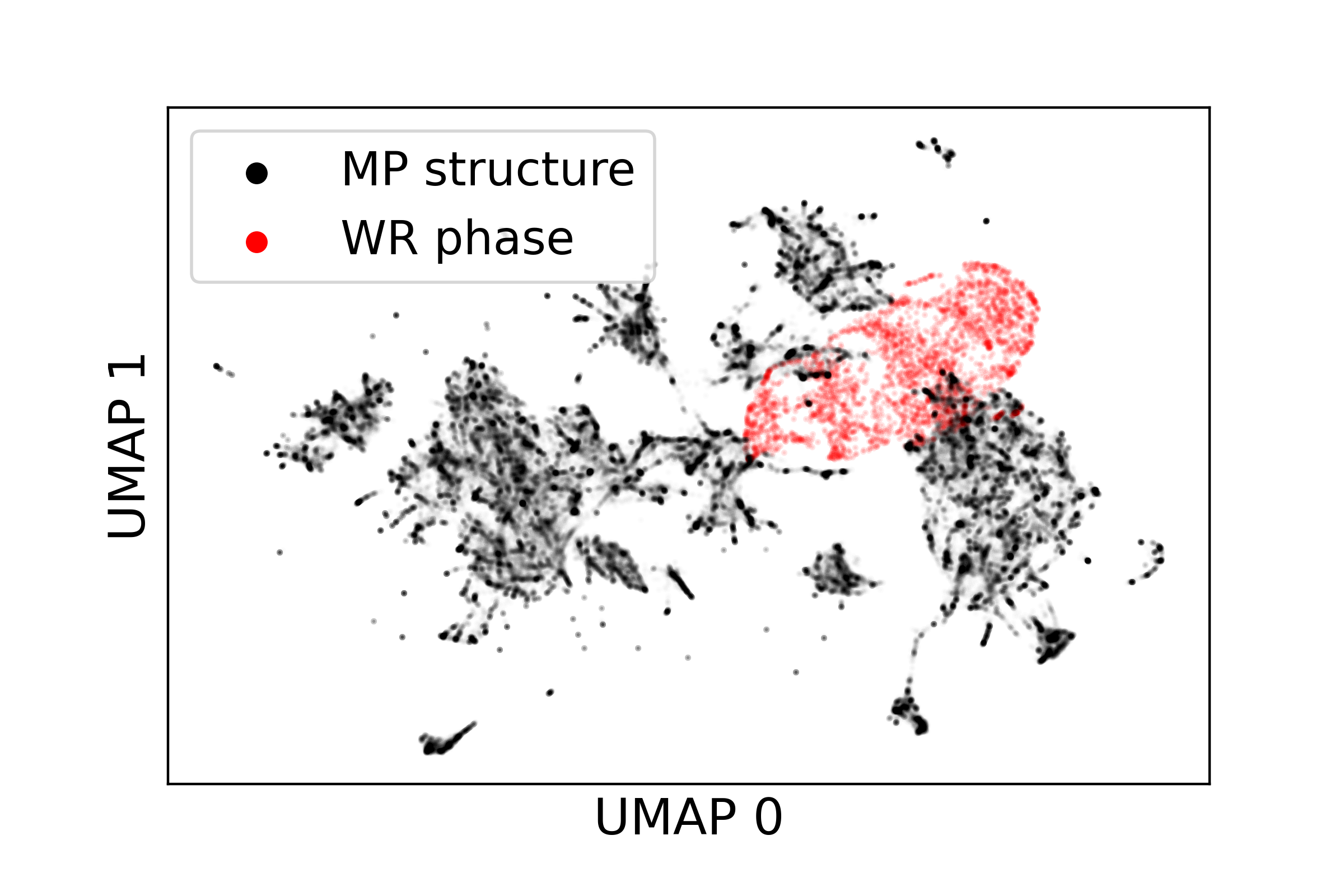}    
    \includegraphics[width=8cm]{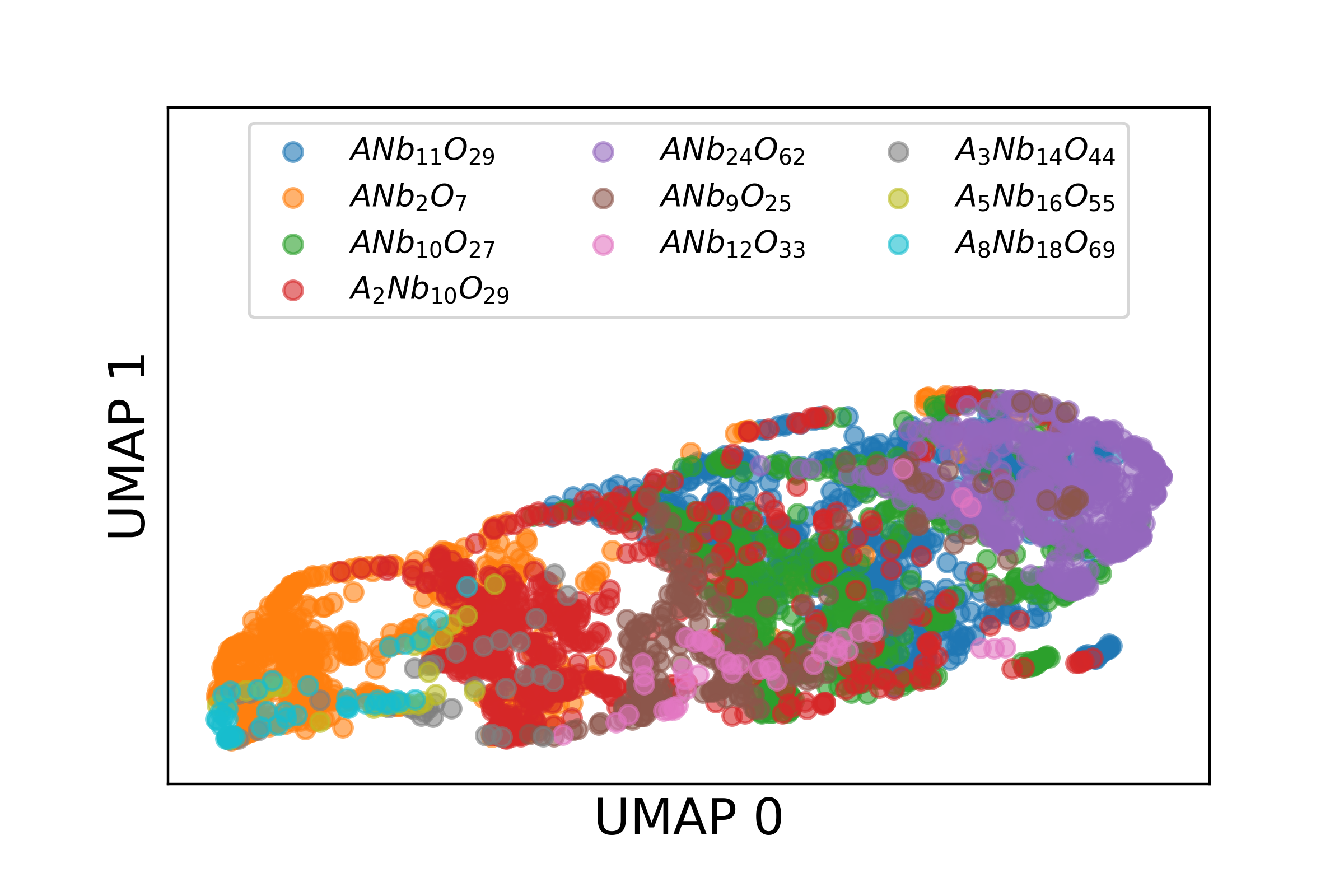}
    \caption{(Left panel) UMAP analysis of the MP database compared with our database of WR phases. (Right panel) Comparison of the 3283 compositions across the 10 prototypes.}
    \label{fig:umap}
\end{figure}

Very few of the WR phase niobate compositions reported in this work have been previously analyzed either experimentally or theoretically. In fact, very few of these structure types can be found in the Materials Project (MP) database~\cite{jain2013commentary}. Left panel of Figure \ref{fig:umap}, compares the WR phase structures computed in this work to all structures in the MP database after applying the dimensionality reduction algorithm Uniform Manifold Approximation Projection (UMAP) plot~\cite{mcinnes2018umap} using MACE-MP0 features~\cite{batatia2023foundation}. In this case, the high-dimensional MACE-MP0 atomic-centered features are averaged over the structure and then projected to 2D using a UMAP, with structures from the MP database in black and structures computed in this work shown in red. There is little overlap between the red and black points, indicating that the vast majority of the WR phases computed in this work are dissimilar to structures tabulated in the MP and are therefore missing from the MP database. Thus, this work fills a gap of undiscovered WR-niobates. Right panel of Figure \ref{fig:umap} shows the UMAP plot for the WR phases, where each color represents a different prototype. These prototypes are grouped as the major difference between the structures of compositions of the same prototype comes from the DFT-relaxation of the structure.  This data is intended to accelerate the discovery of undiscovered  WR phases by guiding experimental synthesis efforts toward compounds with improved stability. 

\begin{figure}[h!]
    \centering
    \begin{subfigure}[b]{0.49\textwidth}
        \centering
        \includegraphics[width=\textwidth]{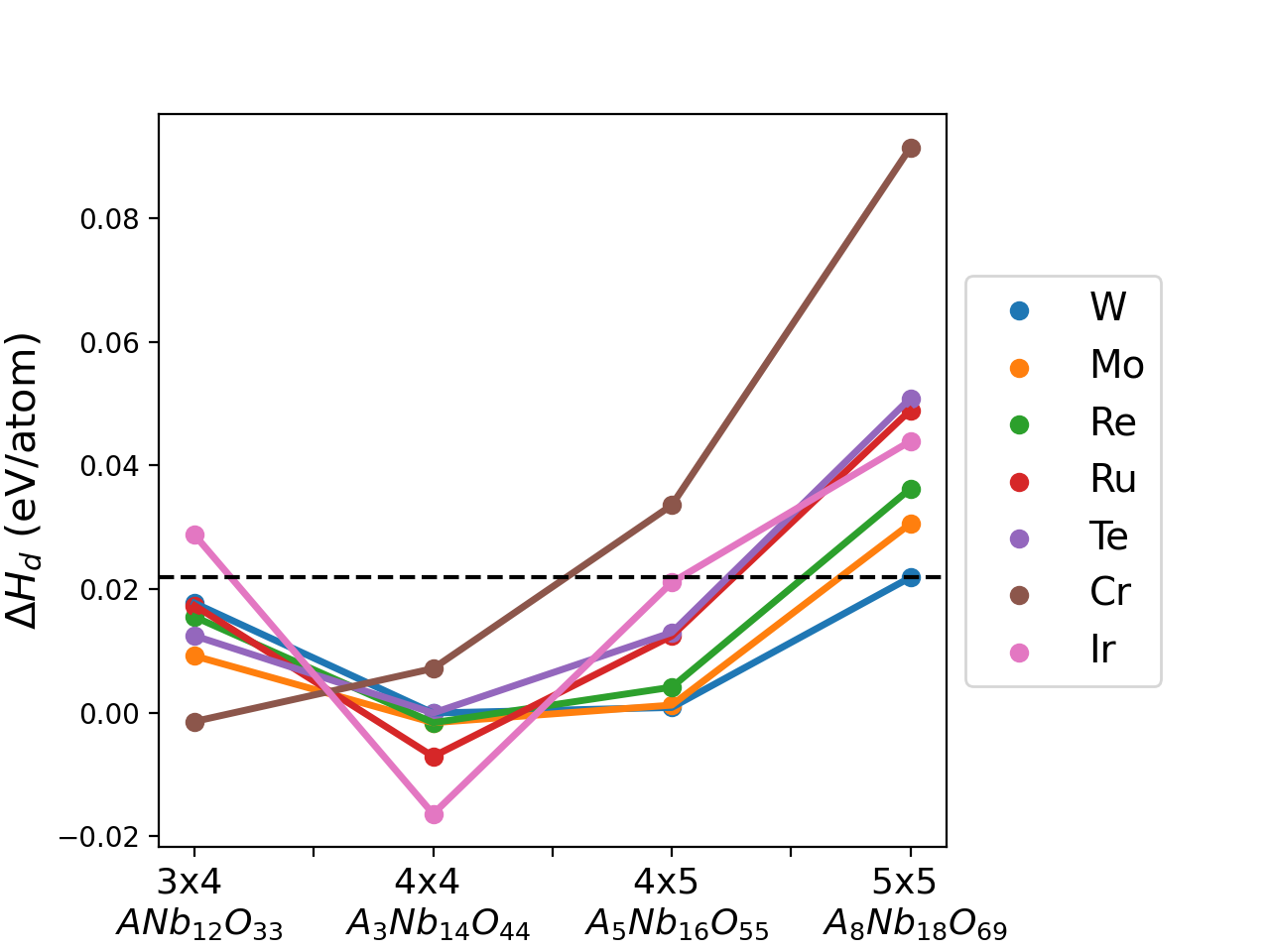}
        \caption{}
        \label{fig:6_block_a}
    \end{subfigure}
    \hfill
    \begin{subfigure}[b]{0.49\textwidth}
        \centering
        \includegraphics[width=\textwidth]{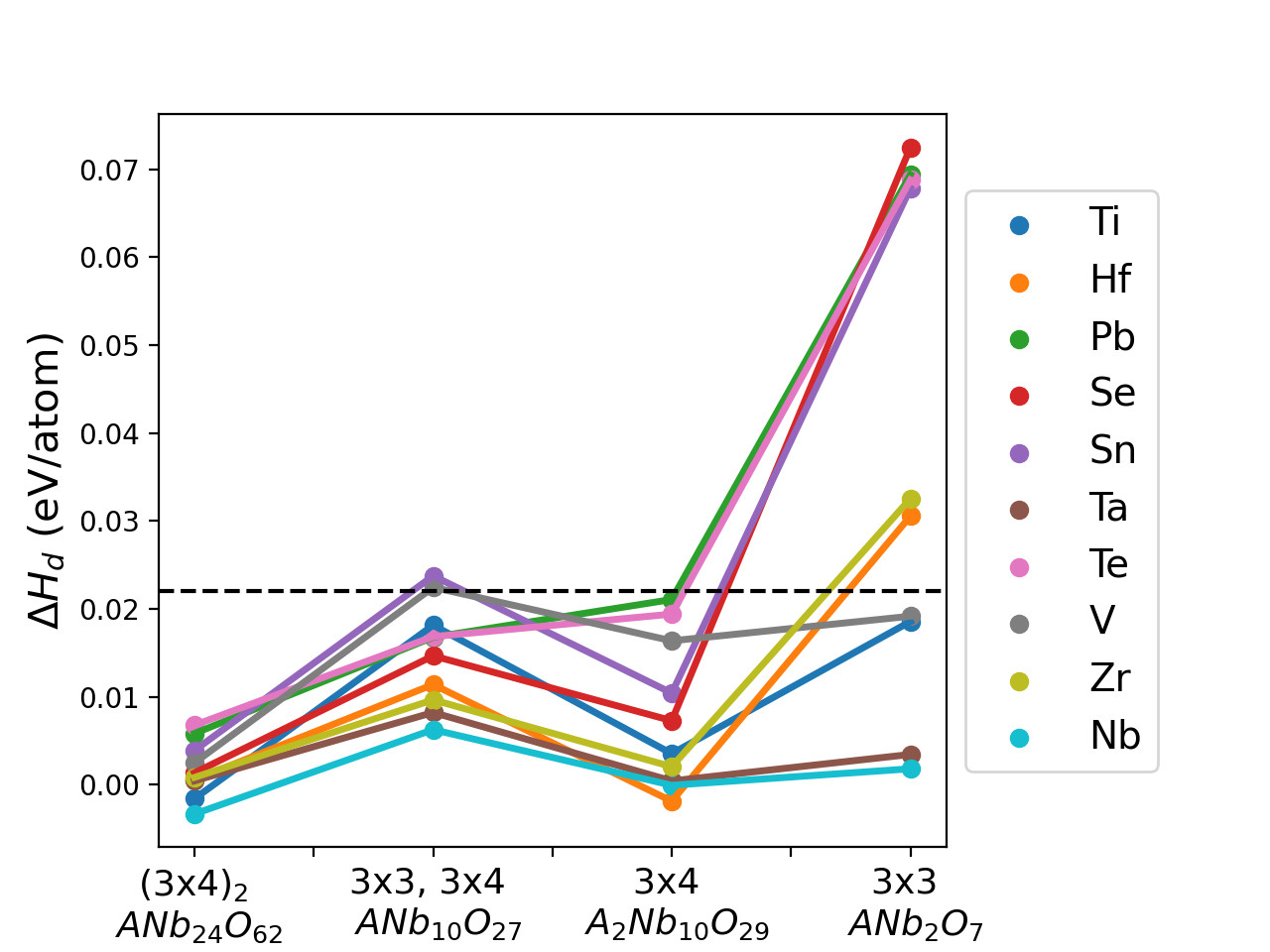}
        \caption{}
        \label{fig:6_block_b}
    \end{subfigure}
    \caption{ \dhd{} as a function of $A$-site element substitution ordered from left to right by increasing block size for prototypes with (a) {\nsix} and  (b) \nfour. Only the 7 {\nsix} and 10 {\nfour} most stable substitutions are shown. The threshold for (meta)stable is shown as a dashed line (unstable above the line).}
    \label{fig:6_block}
\end{figure}

In the following sections, the trends in element substitutions are explored for each $A$-site oxidation state separately to provide a clear understanding of how element substitutions affect stability (see Figure \ref{fig:6_block}). First, for the prototype phases with \nsix, the major difference between the prototypes is the size of the block, which increases with more oxygen per metal (i.e. as block size increases, the oxygen stoichiometry diverges from the O/M ratio of 2.54 for $O_{O33}^{A6+}$ $<$ $O_{O44}^{A6+}$ $<$ $O_{O55}^{A6+}$ $<$ $O_{O69}^{A6+}$).
The change in stability as a function of block size is shown in Figure \ref{fig:6_block}a. In general, the $O_{O69}^{A6+}$ compounds tend to be the least stable because of fewer edge-sharing octahedra. This characteristic and its higher instability may also explain why fewer compositions have been reported for the largest $5\times5$ block size than the rest.

In these structures, there is one tetrahedrally coordinated $A$-site ($A^{tet}$) located at the block corner (see SI Figure S2), and the remaining A-sites are octahedrally coordinated ($A^{oct}$). Therefore, as the block size increases, the ratio $A^{oct}/A^{tet}$ decreases because more A-sites occupy octahedral sites. As the block size increases, Cr$^{6+}$ substitutions become increasingly unfavorable, indicating a preference for tetrahedral coordination. This agrees with trends reported for oxides in the Inorganic Crystal Structure Database (ICSD), where Cr$^{6+}$ prefers 4-fold coordination.\cite{waroquiers2017statistical} The opposite trend is observed for Ir$^{6+}$ and Ru$^{6+}$ substitutions, both of which become more stable with $A^{oct}$ occupation at $O_{44}^{A6+}$ before stability decreases at the largest block sizes. For the cations Mo$^{6+}$, Te$^{6+}$, Re$^{6+}$, and W$^{6+}$, there seems to be minimal preference for $A^{tet}$ vs $A^{oct}$ occupation.

Next, the trends are explored for the prototypes with \nfour. A set of the 10 most stable element substitutions for the prototype phases with \nfour{} are shown in Figure \ref{fig:6_block}b. For $O_{62}^{A4+}$, all of the single substitutions considered, except \ce{PNb24O62} (\dhd{} = 22.3 meV/atom), are (meta)-stable within the 22 meV/atom cutoff energy. This is likely due to the low concentration of $A$-site elements with respect to Nb (1/24) and O (1/62), for which the $A$-site substitution has little effect on the overall stability of the compositions. The effect of A-site substitution on \dhd{} becomes more evident with larger $A$-site stoichiometric ratios. For most element substitutions the $O_{27}^{A4+}$ prototype tends to be less stable than the $O_{29}^{A4+}$ prototype with $O_{7}^{A4+}$ being the least stable. The only exceptions are for Hf, Zr, and Ti which are most stable in the \ce{A2Nb10O29} prototype, and Nb and Ta, which stay consistently stable across all compositions.

Within this set, except for \ce{Hf2Nb10O29}, all $Nb_xO_y$ compounds (i.e., where all $A$-sites are occupied with Nb) tend to be more stable than $A^{+4}_xNb_yO_z$. This trend extends beyond \nfour{} to the prototypes with \nthree{} or \nfive, where Nb tends to be one of the most stable $A$-site substitutions (see SI Figure S3). For the $O_{29}^{A3+}$ prototype with \nthree, there are 14 out of a total of 24 compounds that are (meta)-stable. These are (ordered from most to least stable): Nb, Ti, Fe, V, Mn, Mo, In, Ga, Cr, Co, Ge, Y, As, Ni, Al, Bi, and Ru, of which compounds have been previously synthesized for 6 (Nb, Fe, Ga, Cr, Co, and Al). For the $O_{25}^{A5+}$ prototype with \nfive, there are 9 out of a total of 14 (meta)-stable compositions. In order of most to least stable, they are: As, Nb, V, P, Ta, Sb, W, Te, and Mo. Interestingly, the 4 most stable of these substitutions (As, Nb, V, and P with \dhd{} values of -7.9, -6.0, -0.05, 3.7 meV/atom, respectively) are compounds that have been previously experimentally synthesized: \ce{AsNb9O25}, \ce{Nb2O5}, \ce{VNb9O25}, and \ce{PNb9O25}. Overall, except for \ce{Hf2Nb10O29}, \ce{AsNb9O25}, and the \nsix{} prototypes (Nb does not adopt a 6+ oxidation state), $Nb_xO_y$ compounds tend to be more stable than $A_xNb_yO_z$ compounds. 

\subsection{Trends in the calculated stability for $A$/$A^{'}$ substituted compositions (double substitutions)}

\begin{figure}[h!]
    \centering
    \includegraphics[width=8cm]{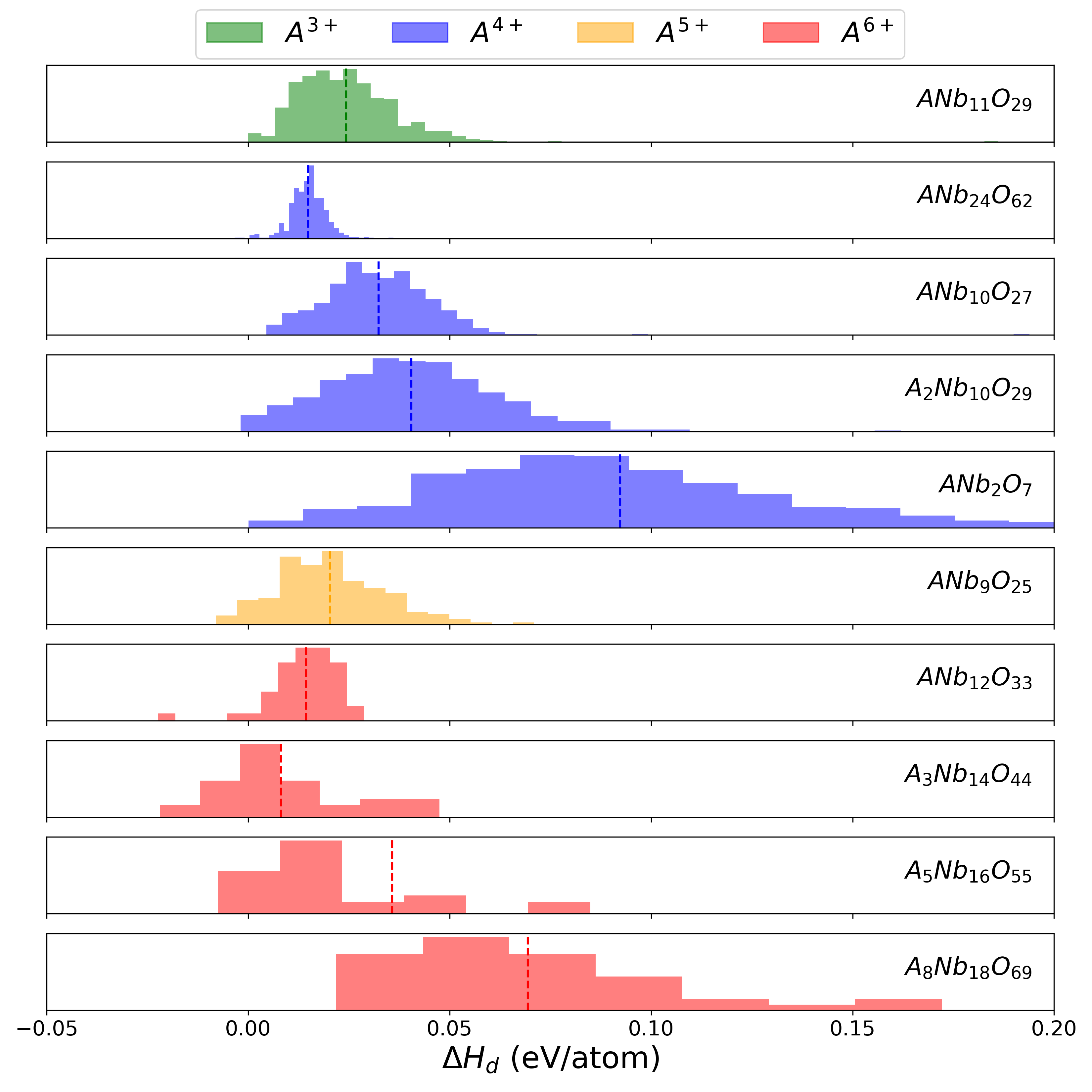}    
    \caption{Distribution of \dhd{} for all double A-site substitutions ($A$/$A^{'}$) separated by prototype and colored by average $A$-site oxidation state. The colored dashed lines show the means of the distributions.
    }
    \label{fig:dhd_hist}
\end{figure}

While single substitutions are limited to the oxidation state of the $A$-site prototype, double-site substitutions can encompass different oxidation states  while maintaining charge balance (e.g. \nfive{} average based on combining equal portions of \nsix{} with $A'^{4+}$), opening the opportunity to identify further combinations in the pursuit of high performance materials. 
Herein double-site substitutions are defined as replacing 50\% of the $A$-sites listed in the general formula with an $A^{+n_{A}}$ cation and the remaining 50\% with a second $A^{'+n_{A^{'}}}$ cation. All combinations of oxidation states that result in a charge-balanced composition were considered for substitutions. For example, the $A$-site cation with \nfour{} can be substituted by the combinations +4/+4, +5/+3, +6/+2, and \nthree{} can be substituted by combinations +3/+3, +4/+2, +5/+1. The distributions of \dhd{} for all double-site substitutions are shown in Figure \ref{fig:dhd_hist}. On average, the most stable substitutions are found for prototypes $O_{44}^{A6+}$, followed by $O_{33}^{A6+}$, $O_{62}^{A4+}$, $O_{25}^{A5+}$, and $O_{29}^{A3+}$. The least stable substitutions are found for prototypes $O_{69}^{A6+}$ and $O_{7}^{A4+}$.

For the prototype phases with \nsix, the distributions of \dhd{} follow a similar trend with respect to block size as the single-site substitutions. Like single-site substitutions, the $4\times4$ block is the most stable followed by the $3\times3$ block, the larger $4\times5$ is less stable, and the largest $5\times5$ is the least stable on average (average \dhd{} follows $O_{O44}^{A6+}$ $<$ $O_{O33}^{A6+}$ $<$ $O_{O55}^{A6+}$ $<$ $O_{O69}^{A6+}$). The only (meta)-stable composition in the $5\times5$ block size is \ce{W8Nb18O69} indicating that cation substitution alone (at least on the $A$-sites) will not improve the stability of the largest block size. Of all the \nsix{} cations considered, Mn substitution decreases stability whereas Re and W tend to improve stability (see SI Figure S4). In the $4\times4$ block ($O_{O44}^{A6+}$), all W substitutions are (meta)-stable within the \dhd{} $<$ 22 meV/atom cutoff energy except when co-substituted with Mn. This includes the experimentally observed compound \ce{Mo$_{1.5}$W$_{1.5}$Nb14O44} which is stable with \dhd{} $=$ -4.7 meV/atom. In this compound, double $A$-site substitution increases the stability relative to the single substitutions \ce{Mo3Nb14O44} and \ce{W3Nb14O44}, see Table S1.

The grid plots for the double cation substitutions (with the original prototype oxidation state of \nfour{}) are shown in the SI Figure S5. Like the single substitutions, the double $A$-site substitutions in $O_{O62}^{A4+}$  have little effect on stability with 95\% of the 558 compounds having \dhd{} $<$ 22 meV/atom (see Figure \ref{fig:dhd_hist}). As the stoichiometric ratio of the $A$-site increases, more compounds become unstable with 20.2\%, 19.5\%, and 3.2\% having \dhd{} $<$ 22 meV/atom for $O_{O27}^{A4+}$, $O_{29}^{A4+}$, and $O_{O7}^{A4+}$, respectively. As shown in Figure \ref{fig:dhd_hist}, the average \dhd{} increases (i.e., less stable) as the ratio of the two elements substituted on the $A$-site increases (average \dhd{} follows $O_{O62}^{A4+}$ $<$ $O_{O27}^{A4+}$ $<$ $O_{29}^{A4+}$ $<$ $O_{O7}^{A4+}$). For the least stable \nfour{} prototype, $O_{O7}^{A4+}$, Ta and Nb substitutions improve stability in addition to some combinations with $A$ = Ti, V, Hf, W, and Zr. Overall, the least stable substitutions for these \nfour{} prototypes are $A$ = Mn, Au, P, and Si. 

For the single element substitutions, we showed that $Nb_xO_y$ (i.e., where all of the $A$ sites are occupied with Nb) compounds tend to be more stable than $A_xNb_yO_z$ compounds. This is also evident for the \nfour{} prototypes where Nb substitution improves stability, and also for the prototype $O_{29}^{A3+}$, where the double A-site substitutions are more stable on average compared to $O_{29}^{A4+}$ compositions which have one less Nb per formula unit. There are 324 $O_{29}^{A3+}$ compositions with \dhd{} $<$ 22 meV/atom, which is the second most behind $O_{O62}^{A4+}$ which has the most stable compositions (532) and the largest Nb content. Moreover, Nb substitutions also result in some of the most stable compositions for the \nfive{} prototype ($O_{25}^{A5+}$), where the most stable compositions are composed of As, Ge, Nb, Si, Ti, V, P, Cr, Ta, Sb, and Zr substitutions, and the least stable compositions come from Ni, Fe, Pd, Ru, Rh, and Ir substitutions. In this prototype oxidation pairs of +4/+6 and +5/+5 were considered for A-site substitutions. On average, the +5/+5 substitutions tend to result in more stable compositions compared to +4/+6, with average \dhd{} of +15 meV/atom and +22 meV/atom above the convex hull, respectively.

For all prototypes, we find that larger differences between oxidation states on the A-sites tend to result in less stable compositions on average (see SI Figure S6). For example, the largest difference between A-site oxidation states is the +1/+5 charge pair which tends to result in unstable compositions. $O_{29}^{A3+}$ is the only prototype considered that can accommodate +1 cations on the A-site with the charge pair +1/+5. However, very few +1 cation substitutions are stable. There are no K$^{1+}$ or Cs$^{1+}$ compounds within the cutoff energy and Rb$^{1+}$ is only stabilized on the A-site when co-substituted with Nb$^{5+}$. Na$^{1+}$ or Ag$^{1+}$ substitutions are meta-stable when co-substituted with either Nb$^{5+}$ or Ta$^{5+}$, and Li$^{1+}$ or Cu$^{1+}$ substitutions are only meta-stable when co-substituted with Nb$^{5+}$, Ta$^{5+}$, or V$^{5+}$. In addition to the instability observed for +1 cations, substitutions of the larger cations Ba$^{2+}$ and Sr$^{2+}$ as well as P$^{4+}$/P$^{5+}$ in the prototype $O_{29}^{A3+}$ also tend to result in unstable compositions.  

\subsection{Experimental synthesis of discovered compound}

\begin{figure}[h!]
    \centering
    \includegraphics[width=10cm]{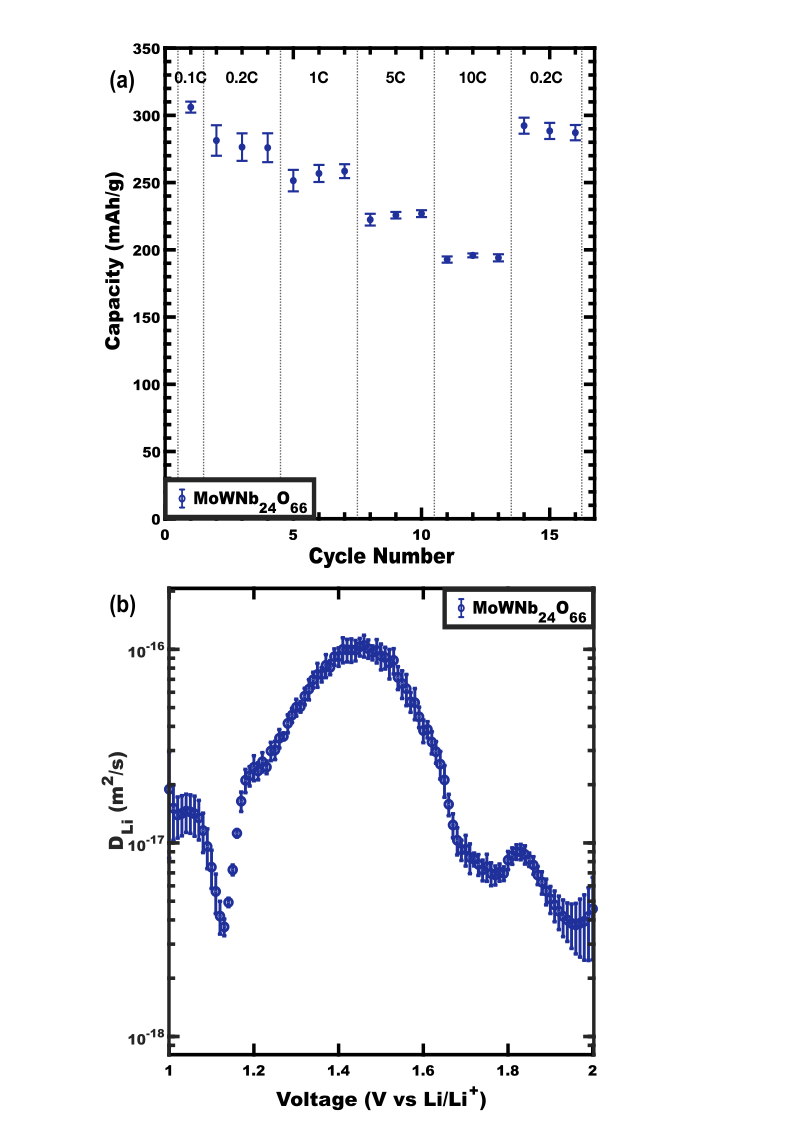}
    \caption{Experimental data on \ce{MoWNb24O66} including the (a) C-rates dependent galvanostatic lithiation capacities and the (b) lithium diffusivities. Mean values shown from three samples including the error-of-the-mean.}
    \label{fig:ecif}
\end{figure}

Our results presented here indicate that prototypes with the largest block size of $5\times5$ ($O_{O69}^{A6+}$) are the least stable whereas smaller block sizes such as 3$\times$4 and 4$\times$4 are the most stable. These small block prototypes were furthermore predicted to have many stable double substitutions which opens up significant composition space. We thus examined the synthesis and performance of a candidate compound  (\ce{Mo$_{0.5}$W$_{0.5}$Nb$_{12}$O$_{33}$}) which was closely related to the \ce{MoNb$_{12}$O$_{33}$} we recently studied.\cite{muhit2024comparison} This target compound is also in a similar chemical space that had previously been investigated (i.e., Mo-W-Nb-O) including cation mixing with a larger (4$\times$4) block size (\ce{Mo$_{1.5}$W$_{1.5}$Nb$_{14}$O$_{44}$} \cite{tao2022insight}).

The computationally predicted phase of \ce{Mo$_{0.5}$W$_{0.5}$Nb$_{12}$O$_{33}$} (the double formula is \ce{MoWNb24O66}) was synthesized and its lithiation behavior was measured. A sol-gel synthesis route was used to prepare the phase where the X-ray diffraction data and Rietveld refinement were consistent with the predicted structure, having overall fit parameter (Rwp) on the order of 10.3\% (Figure S7). The values of the refined lattice parameters (a, b, c, $\alpha$, $\beta$, $\gamma$), lattice volume (V), atom positions (x, y, z) are presented in Tables S2,3. A recent benchmark WR example for \ce{Nb16W5O55} demonstrated C/5 capacities of 225 mAh/g which decreased at higher current densities, e.g. at 5C the capacity was 171 mAh/g~\cite{griffith2018niobium}. Similarly, the 5C capacity for a closely related benchmark compound of \ce{Nb18W8O69} was also $\sim$160-200 mAh/g,~\cite{griffith2020superionic} depending on the heat treatment and voltage window used. For comparison, the \ce{MoWNb24O66} presented for the first time here (Figure \ref{fig:ecif}a) resulted in 278 $\pm$ 2 mAh/g at C/5 (23.6\% higher) and maintained that improved capacity at higher rates, e.g. at 5C the capacity was 225 ± 1 mAh/g (31.6\% higher). Yet faster cycling at 10C (6-minute charge/discharge) maintained a capacity of 194 $\pm$ 1 mAh/g. Subsequently, returning to slower C/5 cycling did not reveal any notable capacity loss as a result of rate testing. No attempts were made to optimize the active material particle size, the electrode additives, or calendaring since this is a proof-of-concept demonstration. From this perspective, fundamental material properties are the focus and thus the diffusivity values were determined using intermittent current interruption (ICI). For comparison, \ce{Nb16W5O55} was reported to have a diffusivity of $1 \cdot 10^{-16}$ m$^2$/s between 1.4-1.6 V vs Li/Li$^+$~\cite{salzer2023structure}. The lithium diffusivity in \ce{MoWNb24O66} varied from $3.7 \cdot 10^{-18}$ to $1.0 \cdot 10^{-16}$ m$^2$/s with the peak value at 1.45 V vs Li/Li$^+$. Capacity-weighted figures-of-merit enable the comparison of material transport characteristics in a way that includes a wide-range of lithiation extents to provide a more comprehensive evaluation than single-point metrics.~\cite{sturgill2025capacity} The capacity-weighted diffusivity for \ce{MoWNb24O66} was determined to be $2.9 \cdot 10^{-17}$ ±  $2.7 \cdot 10^{-18}$ m$^2$/s. Thus a computationally predicted phase was realized experimentally with performance comparable to and somewhat exceeding a recent WR benchmark for \ce{Nb16W5O55}.

\section{Conclusions}

This study has systematically explored the trends in single and double A-site substitutions for the 10 WR-niobate prototype structures, providing comprehensive insights into their stability. By utilizing DFT calculations, we evaluated the stability (\dhd{}) of 3283 potential compositions, identifying 1301 (meta)-stable compositions with \dhd{} $<$ 22 meV/atom, which is considered to be within the experimentally viable stability range. Our findings highlight that many of these compositions have not been previously reported, positioning them as out of distribution compared to existing materials in the MP database. This underscores the novelty of the WR-niobate compositions identified in this work. The study reveals that block size and coordination preferences significantly influence stability, with larger block sizes generally resulting in less stable compositions. Double-site substitutions show that maintaining similar oxidation states on the A-sites can enhance stability and that higher Nb content results in more stable compositions. 
A predicted WR compound Mo$_{0.5}$W$_{0.5}$Nb$_{12}$O$_{33}$ was synthesized where its diffusivity and C-rate performance exceeded the recent benchmark material \ce{Nb16W5O55}. Overall, this work provides a detailed map of stability trends that can guide experimental synthesis efforts, accelerating the discovery of new niobate materials with enhanced properties.

\section{Methods}
\subsection{DFT Calculations}
GGA+U DFT calculations were performed using the Vienna Ab-initio Simulation Program (VASP)\cite{kresse1993ab,kresse1996efficiency,kresse1996efficient} with the Perdew-Burke-Ernzerhof (PBE)\cite{perdew1996generalized} exchange-correlation functional and periodic boundary conditions utilizing projector augmented wave (PAW)\cite{kresse1994norm,kresse1999ultrasoft} pseudopotentials. All calculations are compatible with the legacy version of the MP database\cite{jain2013commentary,ong2015materials}, which tabulates the structures and energies of inorganic materials computed at the PBE (GGA+U) level of theory. The electronic wave functions were expanded in a plane wave basis set with an energy cutoff of 520 eV. The Brillouin zones were sampled during geometry optimizations using the Monkhorst-Pack algorithm to automatically generate a $\Gamma$-point centered k-point mesh with a grid density of at least 1000 $\times$ atoms $^{-1}$ $\times$ unit cell$^{-1}$. All geometry optimizations were initialized with a ferromagnetic spin configuration and performed without symmetry constraints on the optimized wavefunction. Lattice vectors and internal coordinates were optimized so that total energies converged to within $10^{-6}$ eV and forces converged to within 0.01 eV/\AA. The pseudopotentials used are consistent with pymatgen’s MPRelaxSet. Hubbard U parameters for the elements Mn, Fe, Co, Cr, Mo, W, V, and Ni were taken from the Pymatgen\cite{ong2013python} python package’s MPRelaxSet, which tabulates U parameters calibrated using the approach described by Wang et al.\cite{wang2006oxidation} Energy corrections were applied to each composition using Pymatgen's MaterialsProject2020Compatibility. \cite{wang2021framework}

\subsection {Prototype Structures and Element Substitution}
 
\begin{figure}[h!]
    \centering
    \includegraphics[width=8cm]{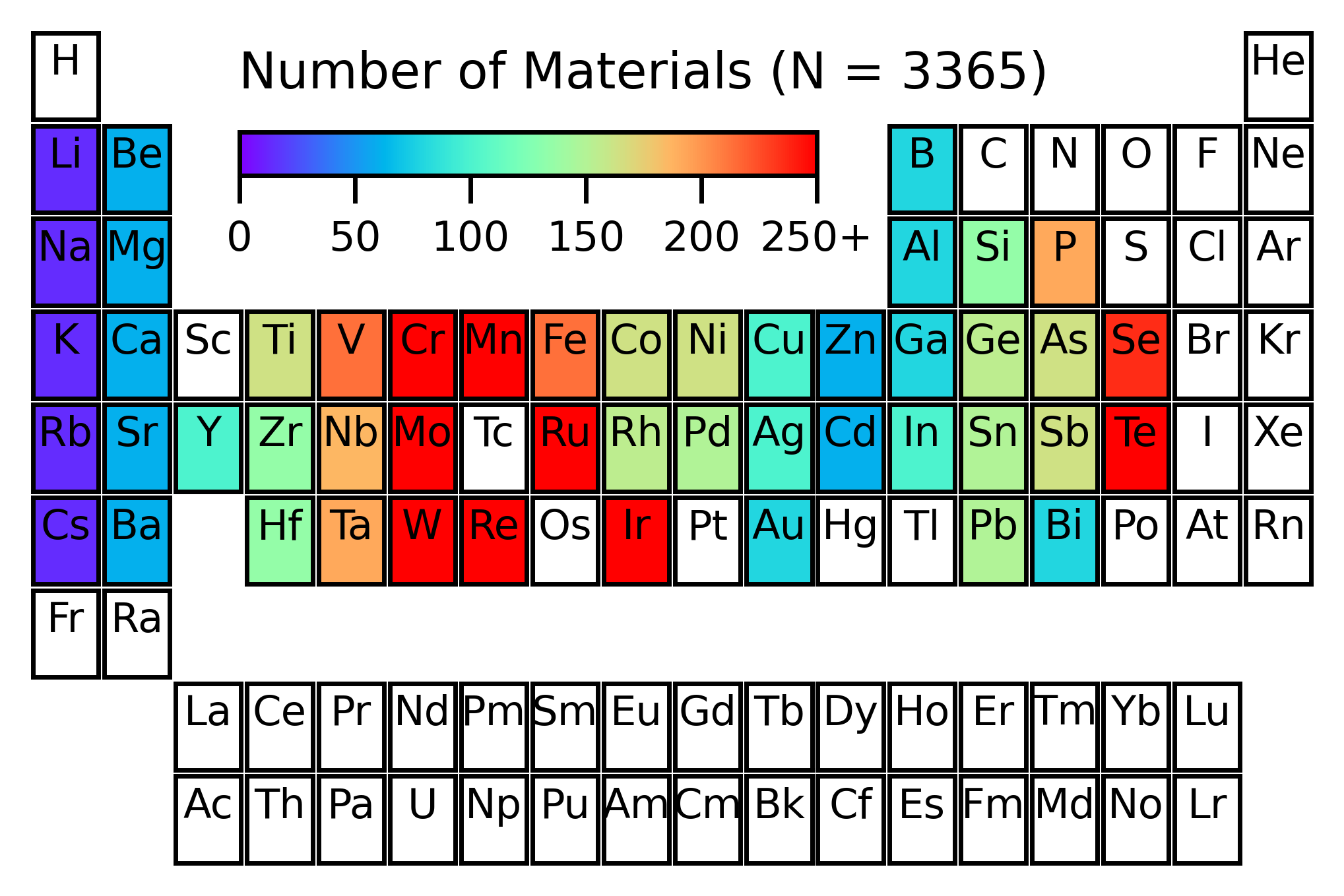}    
    
    \caption{Periodic table plot of all single and double A-site substitutions considered in this work.}
    \label{fig:periodic}
\end{figure}

\begin{table*}
\caption{ Description of the 10 experimentally observed WR niobate shear structure prototype phases considered in this work. The general formula is listed. The number of compositions are listed for both single ($A$) and double ($A$/$A^{'}$) substitutions. The number of possible $A$ and $A$/$A^{'}$ substitutions are limited by the oxidation state of the $A$-site cation, $N_A$.}

\begin{tabular}{c c c c c c c c}
Prototype & General Formula & Space group & Prototype ID & $A$ & $A$/$A^{'}$ & $A^{n+}$  & Block size \\   
\hline
\ce{FeNb11O29} & \ce{AB11O29}  & Cmc$2_1$ & O$_{29}^{A3+}$ & 24 & 737 & +3 & $3\times4$ \\
\hline
 \ce{TiNb2O7}  & \ce{AB2O7} & Cm & $O_{O7}^{A4+}$ & 25 & 561 & +4 & $3\times3$  \\
 \ce{Nb11O27} & \ce{AB10O27} & P2 & $O_{O27}^{A4+}$ & 25 & 561 & +4 & $3\times3$, $3\times4$ \\
 \ce{Ti2Nb10O29} & \ce{A2B10O29} & C2/m  & O$_{29}^{A4+}$ & 25 & 561 & +4 & $3\times4$ \\
 \ce{TiNb24O62} & \ce{AB24O62} & P1 & $O_{O62}^{A4+}$ & 25 & 561 & +4 & $(3\times4)_{2}$\\
\hline
 \ce{PNb9O25} & \ce{AB9O25} & P4/n  & $O_{O25}^{A5+}$ & 14 & 231 & +5 & $3\times3$ \\
\hline
 \ce{WNb12O33} & \ce{AB12O33} & C2 & $O_{O33}^{A6+}$ & 9 & 45 & +6 & $3\times4$ \\
 \ce{W3Nb14O44} & \ce{A3B14O44} & C2 & $O_{O44}^{A6+}$  & 9 & 4  & +6 & $4\times4$\\
 \ce{W5Nb16O55} & \ce{A5B16O55} & C2 & $O_{O55}^{A6+}$  & 9 & 45 & +6  & $4\times5$ \\
 \ce{W8Nb18O69} & \ce{A8B18O69} & P1 & $O_{O69}^{A6+}$  & 9 & 45  & +6 & $5\times5$ \\
 \end{tabular}
 \label{tab:ptype}
\end{table*}

This work uses 10 experimentally observed WR-niobate shear structures as prototype phases (see Figure \ref{fig:st_im}) for which elements are substituted to identify new stable compositions. Due to charge constraints, 7, 22, 24, 25, 14, and 9 unique elements were considered for the +1, +2, +3, +4, +5, and +6 A-site oxidation states are shown in the SI Figure S1. In total, 48 elements are considered for A-site substitutions (see Figure \ref{fig:periodic}). These cations were chosen based on the likelihood of appearing in tetrahedral or octahedral coordination environments cataloged from $\sim$ 5,000 oxides in the ICSD\cite{waroquiers2017statistical}.

The \ce{FeNb11O29} prototype structure was reported in the Amma space group with partially occupied Nb/Fe sites \cite{pinus2014neutron}. The \ce{TiNb24O62} prototype was reported in the C2 space group with partially occupied Nb/Ti sites.\cite{roth1965mixed} The \ce{Ti2Nb10O29} prototype was reported in the A2/m space group with partially occupied Nb/Ti sites \cite{wadsley1961mixed}. The \ce{W8Nb18O69} prototype was both reported in the I-4 space group with partially occupied Nb/W sites. \cite{roth1965multiple} For these experimentally reported structures with partial cation orderings, the cation ordering with the minimum Ewald energy was used. 

For the remaining prototypes, the cation ordering was used from previous work. The minimum energy cation ordering reported by Ko{\c{c}}er et. al was used for \ce{W3Nb14O44} and \ce{W5Nb16O55}.\cite{koccer2019cation} The \ce{Nb11O27} (\ce{Nb22O54}) prototype was reported in the P2 space group with fully occupied sites.\cite{koccer2019first} The monoclinic \ce{TiNb2O7}, tetragonal \ce{PNb9O25}, and monoclinic \ce{WNb12O33} prototypes were taken from the MP database: mp-1216971, mp-758233, and mp-1198790, respectively, also with fully occupied sites. 

The role of partial cation ordering could be potentially important for the possible synthesis of the compounds, especially considering that materials in the TiO$_2$-Nb$_2$O$_5$ are proposed to be entropy-stabilised on the basis of oxide melt solution calorimetry.~\cite{voskanyan2020entropy} To examine the same, we determined the energy differences between the fully disordered cation configurations—modeled using special quasirandom structures (sqs)~\cite{zunger1990special,van2013efficient}—and the ordered configurations (Table S4). We found that that in all the examined nine samples, the ordered cation configurations are energetically more favored. Moreover, the entropy contributions were small, indicating the cations prefer to remain in an ordered state.
 
A primitive unit cell was constructed and used for single-element substitutions for each prototype to reduce computational expense. However, some primitive unit cells were too small to accommodate double-element substitutions. For these prototypes (\ce{TiNb24O62}, \ce{Nb12WO33}, \ce{Nb14W3O44}, and \ce{Nb16W5O55}), a supercell was constructed by doubling the unit cell in the direction of the smallest lattice vector. Then, the cation ordering for each prototype was determined by a minimum Ewald summation for each unique pair of oxidation states. After element substitution, the prototype volumes were scaled using the Data-Mined Lattice Scaling (DLS) \cite{hautier_data_2011} method, which estimates crystal volume based on data-mined bond lengths, to generate a better initial structure of each composition prior to optimization with DFT. 

\subsection{Phase Stability}
 
\begin{figure}[h!]
    \centering
    \includegraphics[width=8cm]{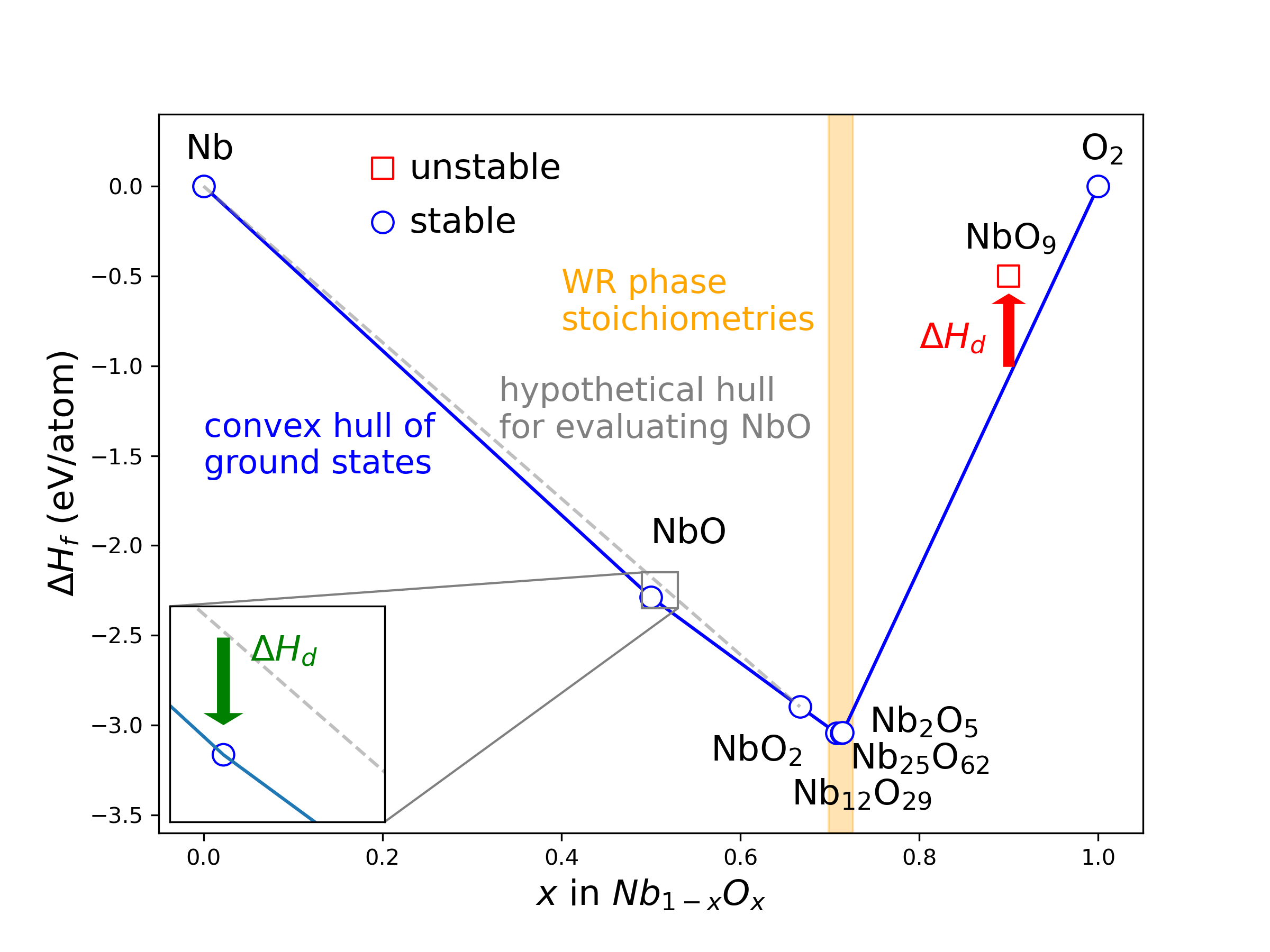}    
    \caption{Illustration of the convex hull construction for the Nb-O binary chemical system. All phases shown are tabulated in the MP database except \ce{Nb25O62}, which is a WR phase calculated in this work, and \ce{NbO9}, which is a fictitious phase used only to illustrate the calculation of \dhd{}.}
    \label{fig:chull}
\end{figure}

The convex hull is a geometric concept used to determine the stability of a set of compounds. For a given chemical system, the convex hull is the smallest convex set that contains all the energy points of the system's compounds. A compound is stable if its energy lies on the convex hull of ground states, see the blue line and blue open circles in Figure \ref{fig:chull}. Otherwise, a compound is metastable or unstable, see the red square in Figure \ref{fig:chull}. To construct a convex hull, the formation energies of all possible compounds in the chemical system are plotted as a function of composition. The formation energy, \dhf, is the energy difference between a phase and its constituent elements. In general, for a phase composed of $N$ elements, \dhf $ = \sum_{i=1}^{N} \mu_{i} n_{i}$, where $n_{i}$ is the total number of element $i$ and $\mu_{i}$ is the elemental reference for element $i$.

The magnitude of (in)-stability is given by the decomposition enthalpy (\dhd{}). For compounds that lie above the convex hull, \dhd{} is the energy difference between that compound and the energy of the convex hull at the same composition, see red arrow in Figure \ref{fig:chull}. For stable compounds that lie on the convex hull, a hypothetical hull is constructed (see gray dashed line in Figure \ref{fig:chull}) by removing the compound of interest, and \dhd{} is computed as the energy difference between that compound and the energy of the hypothetical hull at the same composition, see green arrow in the inset axis in Figure \ref{fig:chull}. In this work, the convex hull construction utilizes all competing phases computed herein as well as those tabulated in the MP database. For the binary Nb-O chemical system, all phases shown in Figure \ref{fig:chull} are tabulated in the MP database except \ce{Nb25O62}, which is a WR phase calculated in this work, and \ce{NbO9}, which is a fictitious phase used only to illustrate the calculation of \dhd{}. In the binary Nb-O chemical system, the WR niobates (\ce{Nb12O29}, \ce{Nb25O62}, and \ce{Nb2O5}) have the most negative  \dhf{} values and are therefore found at the bottom of the convex hull. Because Nb is typically found in the +5 oxidation state on the B-site of the WR phases (see the general formulae in Table \ref{tab:ptype}), all WR prototype phases considered in this work lie within a small compositional window, see the orange bar in Figure \ref{fig:chull}.

\subsection{Experimental Synthesis}
\ce{MoWNb24O66} phase was synthesized via a sol-gel route. Specifically, 0.465 mL of Niobium (V) ethoxide (Sigma Aldrich, 99.9\%), 30.1 mg of Tungsten (VI) chloride (Thermo Scientific, 99\%), and 24.7 mg of Molybdenum(VI) dioxide bis(acetylacetonate) (Strem, $>$95\%) were combined in a solution of 4.0 mL of ethanol (Decon Labs, 100\%) and 25 $\mu L$ of nitric acid (Fisher Chemical, 66\%) with overall molar ratios of the metal precursors being 24:1:1, respectively. The solution was stirred briefly, cast in a Teflon dish on a hot plate set to 105 $\degree$C until yielding a dry powder. The powder was heated in a furnace to 900  $\degree$C for 6 h in air. Initial wide-angle X-ray scattering data were acquired at the South Carolina SAXS Collaborative. More detailed X-ray diffraction (XRD) analysis of \ce{MoWNb24O66} was conducted on a lightly ground powder loaded into a 0.3 mm diameter Kapton capillary. The sample was measured using a D8 Advance diffractometer (Bruker AXS) with Mo K$\alpha_1$ radiation ($\lambda$ = 0.7093 \AA) in transmission mode with an incident beam focusing monochromator. The collected data were refined via the Rietveld refinement approach in Topas software (v.7) using the low-symmetry triclinic (P1) computational model for the \ce{MoWNb24O66} WR phase. The refinements employed fundamental parameters peak shape models to account for the instrument geometry and microstructural features of the specimen including crystallite size and crystallite microstrain. It was unnecessary to refine the atom positions from those derived computationally; only the isotropic displacement parameters were added to the computational model as typical values for oxides. Standard galvanostatic charge/discharge measurements were conducted using lithium half-cells from 1.0-3.0 V vs Li/Li$^+$ in a 2032-coin format as reported in detail elsewhere~\cite{muhit2024comparison}. The current density used for each C-rate was calculated by assuming one electron/Li+ per transition metal (1C current density of 195.4mA/g). The specific surface area (0.94 $\pm$ 0.02 m$^2$/g) was calculated from Porod analysis of small-angle X-ray scattering data as described in detail elsewhere~\cite{sturgill2025capacity}. ICI was used to derive diffusivity values~\cite{chien2023rapid}. These measurements were repeated with three samples to determine the reported values (mean and standard error-of-the-mean) for capacity and diffusivity.

\begin{acknowledgement}
All authors acknowledge DOE support (DE-SC0023377). This work made use of the South Carolina SAXS Collaborative.
This research used resources of the National Energy Research Scientific Computing Center (NERSC), a Department of Energy Office of Science User Facility using NERSC award ERCAP0029139 and using computational resources sponsored by the Department of Energy's Office of Energy Efficiency and Renewable Energy and located at the National Renewable Energy Laboratory.
\end{acknowledgement}

\begin{suppinfo}
Periodic plot of element substitutions considered for each oxidation state, experimentally observed compounds ordered by DFT calculated $\Delta H_{d}$ values, A$^{6+}$ prototype structures, double A-site substitutions grid plots for prototypes with $n_{A} = +3$ and $+5$, $A$/$A^{'}$-site substitutions grid plots for prototypes with $n_{A} = +6$ and $n_{A} = +4$, , plot showing the difference in $A$-site oxidation state vs the change in average stability (w.r.t. the most stable $A$-site oxidation pair) for double $A$-site substitutions, the wide-angle X-ray scattering data from \ce{MoWNb24O66}, \ce{MoWNb24O66} lattice parameters and atom positions from Rietveld refinement
\end{suppinfo}

\setlength{\fboxrule}{0 pt}
\begin{tocentry}
    \centering
    \includegraphics[width=15cm]{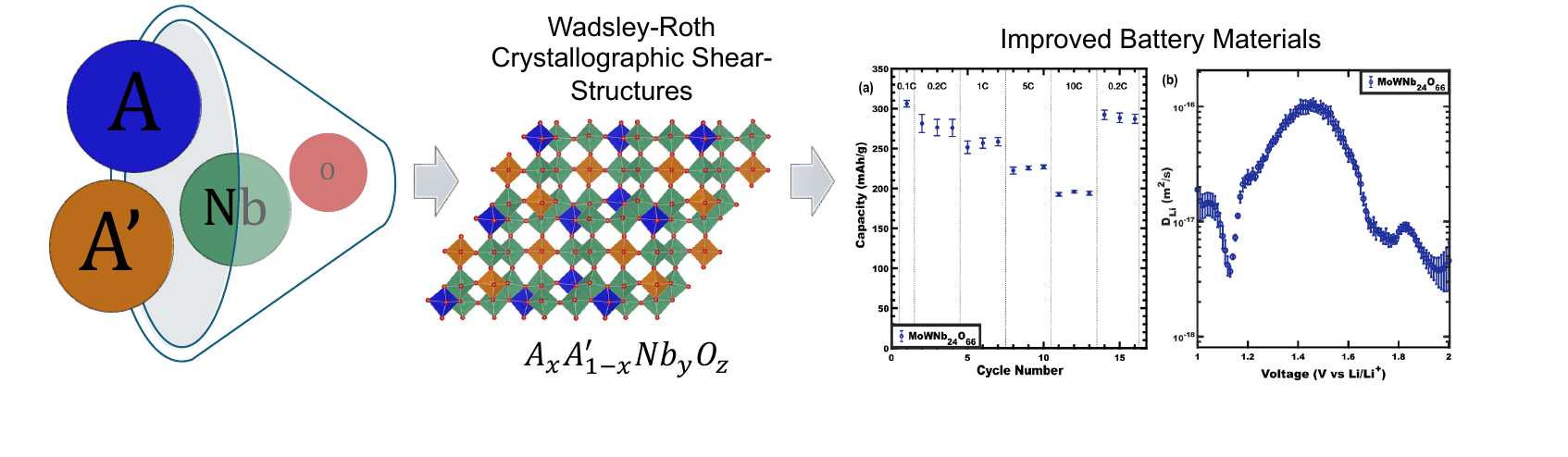}    
\end{tocentry}

\bibliography{main.bib}

\providecommand{\latin}[1]{#1}
\makeatletter
\providecommand{\doi}
  {\begingroup\let\do\@makeother\dospecials
  \catcode`\{=1 \catcode`\}=2 \doi@aux}
\providecommand{\doi@aux}[1]{\endgroup\texttt{#1}}
\makeatother
\providecommand*\mcitethebibliography{\thebibliography}
\csname @ifundefined\endcsname{endmcitethebibliography}  {\let\endmcitethebibliography\endthebibliography}{}
\begin{mcitethebibliography}{65}
\providecommand*\natexlab[1]{#1}
\providecommand*\mciteSetBstSublistMode[1]{}
\providecommand*\mciteSetBstMaxWidthForm[2]{}
\providecommand*\mciteBstWouldAddEndPuncttrue
  {\def\EndOfBibitem{\unskip.}}
\providecommand*\mciteBstWouldAddEndPunctfalse
  {\let\EndOfBibitem\relax}
\providecommand*\mciteSetBstMidEndSepPunct[3]{}
\providecommand*\mciteSetBstSublistLabelBeginEnd[3]{}
\providecommand*\EndOfBibitem{}
\mciteSetBstSublistMode{f}
\mciteSetBstMaxWidthForm{subitem}{(\alph{mcitesubitemcount})}
\mciteSetBstSublistLabelBeginEnd
  {\mcitemaxwidthsubitemform\space}
  {\relax}
  {\relax}

\bibitem[Yang and Zhao(2021)Yang, and Zhao]{yang2021wadsley}
Yang,~Y.; Zhao,~J. Wadsley--Roth crystallographic shear structure niobium-based oxides: Promising anode materials for high-safety lithium-ion batteries. \emph{Advanced Science} \textbf{2021}, \emph{8}, 2004855\relax
\mciteBstWouldAddEndPuncttrue
\mciteSetBstMidEndSepPunct{\mcitedefaultmidpunct}
{\mcitedefaultendpunct}{\mcitedefaultseppunct}\relax
\EndOfBibitem
\bibitem[Liu \latin{et~al.}(2023)Liu, Russo, Montoro, and Pinna]{liu2023recent}
Liu,~Y.; Russo,~P.~A.; Montoro,~L.~A.; Pinna,~N. Recent developments in Nb-based oxides with crystallographic shear structures as anode materials for high-rate lithium-ion energy storage. \emph{Battery Energy} \textbf{2023}, \emph{2}, 20220037\relax
\mciteBstWouldAddEndPuncttrue
\mciteSetBstMidEndSepPunct{\mcitedefaultmidpunct}
{\mcitedefaultendpunct}{\mcitedefaultseppunct}\relax
\EndOfBibitem
\bibitem[Krumeich(2022)]{krumeich2022complex}
Krumeich,~F. The complex crystal chemistry of niobium tungsten oxides. \emph{Chemistry of Materials} \textbf{2022}, \emph{34}, 911--934\relax
\mciteBstWouldAddEndPuncttrue
\mciteSetBstMidEndSepPunct{\mcitedefaultmidpunct}
{\mcitedefaultendpunct}{\mcitedefaultseppunct}\relax
\EndOfBibitem
\bibitem[Ko{\c{c}}er \latin{et~al.}(2019)Ko{\c{c}}er, Griffith, Grey, and Morris]{koccer2019cation}
Ko{\c{c}}er,~C.~P.; Griffith,~K.~J.; Grey,~C.~P.; Morris,~A.~J. Cation disorder and lithium insertion mechanism of Wadsley--Roth crystallographic shear phases from first principles. \emph{Journal of the American Chemical Society} \textbf{2019}, \emph{141}, 15121--15134\relax
\mciteBstWouldAddEndPuncttrue
\mciteSetBstMidEndSepPunct{\mcitedefaultmidpunct}
{\mcitedefaultendpunct}{\mcitedefaultseppunct}\relax
\EndOfBibitem
\bibitem[Cava \latin{et~al.}(1983)Cava, Murphy, and Zahurak]{cava1983lithium}
Cava,~R.~J.; Murphy,~D.~W.; Zahurak,~S. Lithium insertion in Wadsley-Roth phases based on niobium oxide. \emph{Journal of the electrochemical society} \textbf{1983}, \emph{130}, 2345\relax
\mciteBstWouldAddEndPuncttrue
\mciteSetBstMidEndSepPunct{\mcitedefaultmidpunct}
{\mcitedefaultendpunct}{\mcitedefaultseppunct}\relax
\EndOfBibitem
\bibitem[Song \latin{et~al.}(2020)Song, Li, Liu, Zhang, Yan, Tang, Huang, Zhang, and Li]{song2020ultrafast}
Song,~Z.; Li,~H.; Liu,~W.; Zhang,~H.; Yan,~J.; Tang,~Y.; Huang,~J.; Zhang,~H.; Li,~X. Ultrafast and stable Li-(de) intercalation in a large single crystal H-Nb2O5 anode via optimizing the homogeneity of electron and ion transport. \emph{Advanced Materials} \textbf{2020}, \emph{32}, 2001001\relax
\mciteBstWouldAddEndPuncttrue
\mciteSetBstMidEndSepPunct{\mcitedefaultmidpunct}
{\mcitedefaultendpunct}{\mcitedefaultseppunct}\relax
\EndOfBibitem
\bibitem[Li \latin{et~al.}(2019)Li, Zhu, Fu, Liang, Chen, Luo, Dong, Shao, Lin, Wei, \latin{et~al.} others]{li2019nanosheet}
Li,~R.; Zhu,~X.; Fu,~Q.; Liang,~G.; Chen,~Y.; Luo,~L.; Dong,~M.; Shao,~Q.; Lin,~C.; Wei,~R.; others Nanosheet-based Nb 12 O 29 hierarchical microspheres for enhanced lithium storage. \emph{Chemical communications} \textbf{2019}, \emph{55}, 2493--2496\relax
\mciteBstWouldAddEndPuncttrue
\mciteSetBstMidEndSepPunct{\mcitedefaultmidpunct}
{\mcitedefaultendpunct}{\mcitedefaultseppunct}\relax
\EndOfBibitem
\bibitem[Han and Goodenough(2011)Han, and Goodenough]{han20113}
Han,~J.-T.; Goodenough,~J.~B. 3-V full cell performance of anode framework TiNb2O7/spinel LiNi0. 5Mn1. 5O4. \emph{Chemistry of materials} \textbf{2011}, \emph{23}, 3404--3407\relax
\mciteBstWouldAddEndPuncttrue
\mciteSetBstMidEndSepPunct{\mcitedefaultmidpunct}
{\mcitedefaultendpunct}{\mcitedefaultseppunct}\relax
\EndOfBibitem
\bibitem[Griffith \latin{et~al.}(2020)Griffith, Harada, Egusa, Ribas, Monteiro, Von~Dreele, Cheetham, Cava, Grey, and Goodenough]{griffith2020titanium}
Griffith,~K.~J.; Harada,~Y.; Egusa,~S.; Ribas,~R.~M.; Monteiro,~R.~S.; Von~Dreele,~R.~B.; Cheetham,~A.~K.; Cava,~R.~J.; Grey,~C.~P.; Goodenough,~J.~B. Titanium niobium oxide: from discovery to application in fast-charging lithium-ion batteries. \emph{Chemistry of Materials} \textbf{2020}, \emph{33}, 4--18\relax
\mciteBstWouldAddEndPuncttrue
\mciteSetBstMidEndSepPunct{\mcitedefaultmidpunct}
{\mcitedefaultendpunct}{\mcitedefaultseppunct}\relax
\EndOfBibitem
\bibitem[Deng \latin{et~al.}(2020)Deng, Zhu, Liu, Yang, Wang, Shen, Zhang, Wang, Ai, Ren, \latin{et~al.} others]{deng2020synergy}
Deng,~S.; Zhu,~H.; Liu,~B.; Yang,~L.; Wang,~X.; Shen,~S.; Zhang,~Y.; Wang,~J.; Ai,~C.; Ren,~Y.; others Synergy of ion doping and spiral array architecture on Ti2Nb10O29: a new way to achieve high-power electrodes. \emph{Advanced Functional Materials} \textbf{2020}, \emph{30}, 2002665\relax
\mciteBstWouldAddEndPuncttrue
\mciteSetBstMidEndSepPunct{\mcitedefaultmidpunct}
{\mcitedefaultendpunct}{\mcitedefaultseppunct}\relax
\EndOfBibitem
\bibitem[Griffith \latin{et~al.}(2017)Griffith, Senyshyn, and Grey]{griffith2017structural}
Griffith,~K.~J.; Senyshyn,~A.; Grey,~C.~P. Structural stability from crystallographic shear in TiO2--Nb2O5 phases: Cation ordering and lithiation behavior of TiNb24O62. \emph{Inorganic chemistry} \textbf{2017}, \emph{56}, 4002--4010\relax
\mciteBstWouldAddEndPuncttrue
\mciteSetBstMidEndSepPunct{\mcitedefaultmidpunct}
{\mcitedefaultendpunct}{\mcitedefaultseppunct}\relax
\EndOfBibitem
\bibitem[Patoux \latin{et~al.}(2002)Patoux, Dolle, Rousse, and Masquelier]{patoux2002reversible}
Patoux,~S.; Dolle,~M.; Rousse,~G.; Masquelier,~C. A Reversible Lithium Intercalation Process in an ReO3 Type Structure PNb9 O 25. \emph{Journal of the Electrochemical Society} \textbf{2002}, \emph{149}, A391\relax
\mciteBstWouldAddEndPuncttrue
\mciteSetBstMidEndSepPunct{\mcitedefaultmidpunct}
{\mcitedefaultendpunct}{\mcitedefaultseppunct}\relax
\EndOfBibitem
\bibitem[Zheng \latin{et~al.}(2019)Zheng, Qian, Cheng, Yu, Peng, Liu, Zhang, Xia, Zhu, and Shu]{zheng2019fenb11o29}
Zheng,~R.; Qian,~S.; Cheng,~X.; Yu,~H.; Peng,~N.; Liu,~T.; Zhang,~J.; Xia,~M.; Zhu,~H.; Shu,~J. FeNb11O29 nanotubes: Superior electrochemical energy storage performance and operating mechanism. \emph{Nano Energy} \textbf{2019}, \emph{58}, 399--409\relax
\mciteBstWouldAddEndPuncttrue
\mciteSetBstMidEndSepPunct{\mcitedefaultmidpunct}
{\mcitedefaultendpunct}{\mcitedefaultseppunct}\relax
\EndOfBibitem
\bibitem[Spada \latin{et~al.}(2021)Spada, Albini, Galinetto, Versaci, Francia, Bodoardo, Bais, and Bini]{spada2021fenb11o29}
Spada,~D.; Albini,~B.; Galinetto,~P.; Versaci,~D.; Francia,~C.; Bodoardo,~S.; Bais,~G.; Bini,~M. FeNb11O29, anode material for high-power lithium-ion batteries: Pseudocapacitance and symmetrisation unravelled with advanced electrochemical and in situ/operando techniques. \emph{Electrochimica Acta} \textbf{2021}, \emph{393}, 139077\relax
\mciteBstWouldAddEndPuncttrue
\mciteSetBstMidEndSepPunct{\mcitedefaultmidpunct}
{\mcitedefaultendpunct}{\mcitedefaultseppunct}\relax
\EndOfBibitem
\bibitem[Spada \latin{et~al.}(2019)Spada, Quinzeni, and Bini]{spada2019orthorhombic}
Spada,~D.; Quinzeni,~I.; Bini,~M. Orthorhombic and monoclinic modifications of FeNb11O29, as promising anode materials for lithium batteries: Relationships between pseudocapacitive behaviour and structure. \emph{Electrochimica Acta} \textbf{2019}, \emph{296}, 938--944\relax
\mciteBstWouldAddEndPuncttrue
\mciteSetBstMidEndSepPunct{\mcitedefaultmidpunct}
{\mcitedefaultendpunct}{\mcitedefaultseppunct}\relax
\EndOfBibitem
\bibitem[Pinus \latin{et~al.}(2014)Pinus, Catti, Ruffo, Salamone, and Mari]{pinus2014neutron}
Pinus,~I.; Catti,~M.; Ruffo,~R.; Salamone,~M.~M.; Mari,~C.~M. Neutron diffraction and electrochemical study of FeNb11O29/Li11FeNb11O29 for lithium battery anode applications. \emph{Chemistry of Materials} \textbf{2014}, \emph{26}, 2203--2209\relax
\mciteBstWouldAddEndPuncttrue
\mciteSetBstMidEndSepPunct{\mcitedefaultmidpunct}
{\mcitedefaultendpunct}{\mcitedefaultseppunct}\relax
\EndOfBibitem
\bibitem[Griffith and Grey(2020)Griffith, and Grey]{griffith2020superionic}
Griffith,~K.~J.; Grey,~C.~P. Superionic lithium intercalation through 2$\times$ 2 nm2 columns in the crystallographic shear phase Nb18W8O69. \emph{Chemistry of Materials} \textbf{2020}, \emph{32}, 3860--3868\relax
\mciteBstWouldAddEndPuncttrue
\mciteSetBstMidEndSepPunct{\mcitedefaultmidpunct}
{\mcitedefaultendpunct}{\mcitedefaultseppunct}\relax
\EndOfBibitem
\bibitem[Griffith \latin{et~al.}(2018)Griffith, Wiaderek, Cibin, Marbella, and Grey]{griffith2018niobium}
Griffith,~K.~J.; Wiaderek,~K.~M.; Cibin,~G.; Marbella,~L.~E.; Grey,~C.~P. Niobium tungsten oxides for high-rate lithium-ion energy storage. \emph{Nature} \textbf{2018}, \emph{559}, 556--563\relax
\mciteBstWouldAddEndPuncttrue
\mciteSetBstMidEndSepPunct{\mcitedefaultmidpunct}
{\mcitedefaultendpunct}{\mcitedefaultseppunct}\relax
\EndOfBibitem
\bibitem[Ko{\c{c}}er \latin{et~al.}(2020)Ko{\c{c}}er, Griffith, Grey, and Morris]{koccer2020lithium}
Ko{\c{c}}er,~C.~P.; Griffith,~K.~J.; Grey,~C.~P.; Morris,~A.~J. Lithium diffusion in niobium tungsten oxide shear structures. \emph{Chemistry of Materials} \textbf{2020}, \emph{32}, 3980--3989\relax
\mciteBstWouldAddEndPuncttrue
\mciteSetBstMidEndSepPunct{\mcitedefaultmidpunct}
{\mcitedefaultendpunct}{\mcitedefaultseppunct}\relax
\EndOfBibitem
\bibitem[Salzer \latin{et~al.}(2023)Salzer, Diamond, Nieto, Evans, Prieto, and Sambur]{salzer2023structure}
Salzer,~L.~D.; Diamond,~B.; Nieto,~K.; Evans,~R.~C.; Prieto,~A.~L.; Sambur,~J.~B. Structure--property relationships in high-rate anode materials based on niobium tungsten oxide shear structures. \emph{ACS Applied Energy Materials} \textbf{2023}, \emph{6}, 1685--1691\relax
\mciteBstWouldAddEndPuncttrue
\mciteSetBstMidEndSepPunct{\mcitedefaultmidpunct}
{\mcitedefaultendpunct}{\mcitedefaultseppunct}\relax
\EndOfBibitem
\bibitem[Muhit \latin{et~al.}(2024)Muhit, Wechsler, Bare, Sturgill, Keerthisinghe, Grasser, Morrison, Sutton, Stefik, and zur Loye]{muhit2024comparison}
Muhit,~M. A.~A.; Wechsler,~S.~C.; Bare,~Z.~J.; Sturgill,~C.; Keerthisinghe,~N.; Grasser,~M.~A.; Morrison,~G.; Sutton,~C.; Stefik,~M.; zur Loye,~H.-C. Comparison of Lithium Diffusion in Isostructural Ta12MoO33 and Nb12MoO33: Experimental and Computational Insights from Single Crystals. \emph{Chemistry of Materials} \textbf{2024}, \emph{36}, 10626--10639\relax
\mciteBstWouldAddEndPuncttrue
\mciteSetBstMidEndSepPunct{\mcitedefaultmidpunct}
{\mcitedefaultendpunct}{\mcitedefaultseppunct}\relax
\EndOfBibitem
\bibitem[Liu and Chen(2024)Liu, and Chen]{liu2024wadsley}
Liu,~H.; Chen,~C. Wadsley--Roth phase CoNb 11 O 29 as a high-performance anode for lithium-ion batteries. \emph{Journal of Materials Chemistry A} \textbf{2024}, \relax
\mciteBstWouldAddEndPunctfalse
\mciteSetBstMidEndSepPunct{\mcitedefaultmidpunct}
{}{\mcitedefaultseppunct}\relax
\EndOfBibitem
\bibitem[Lou \latin{et~al.}(2019)Lou, Li, Zhu, Luo, Chen, Lin, Li, and Zhao]{lou2019new}
Lou,~X.; Li,~R.; Zhu,~X.; Luo,~L.; Chen,~Y.; Lin,~C.; Li,~H.; Zhao,~X. New anode material for lithium-ion batteries: aluminum niobate (AlNb11O29). \emph{ACS Applied Materials \& Interfaces} \textbf{2019}, \emph{11}, 6089--6096\relax
\mciteBstWouldAddEndPuncttrue
\mciteSetBstMidEndSepPunct{\mcitedefaultmidpunct}
{\mcitedefaultendpunct}{\mcitedefaultseppunct}\relax
\EndOfBibitem
\bibitem[Fu \latin{et~al.}(2018)Fu, Liu, Hou, Pu, Lin, Yang, Zhu, Hu, Lin, Luo, \latin{et~al.} others]{fu2018highly}
Fu,~Q.; Liu,~X.; Hou,~J.; Pu,~Y.; Lin,~C.; Yang,~L.; Zhu,~X.; Hu,~L.; Lin,~S.; Luo,~L.; others Highly conductive CrNb11O29 nanorods for use in high-energy, safe, fast-charging and stable lithium-ion batteries. \emph{Journal of Power Sources} \textbf{2018}, \emph{397}, 231--239\relax
\mciteBstWouldAddEndPuncttrue
\mciteSetBstMidEndSepPunct{\mcitedefaultmidpunct}
{\mcitedefaultendpunct}{\mcitedefaultseppunct}\relax
\EndOfBibitem
\bibitem[Lou \latin{et~al.}(2017)Lou, Fu, Xu, Liu, Lin, Han, Luo, Chen, Fan, and Li]{lou2017ganb11o29}
Lou,~X.; Fu,~Q.; Xu,~J.; Liu,~X.; Lin,~C.; Han,~J.; Luo,~Y.; Chen,~Y.; Fan,~X.; Li,~J. GaNb11O29 nanowebs as high-performance anode materials for lithium-ion batteries. \emph{ACS Applied Nano Materials} \textbf{2017}, \emph{1}, 183--190\relax
\mciteBstWouldAddEndPuncttrue
\mciteSetBstMidEndSepPunct{\mcitedefaultmidpunct}
{\mcitedefaultendpunct}{\mcitedefaultseppunct}\relax
\EndOfBibitem
\bibitem[Liang \latin{et~al.}(2021)Liang, Huang, Lin, Liang, Huang, Chen, Li, Feng, Lin, and Huang]{liang2021micro}
Liang,~M.; Huang,~Y.; Lin,~Y.; Liang,~G.; Huang,~C.; Chen,~L.; Li,~J.; Feng,~Q.; Lin,~C.; Huang,~Z. Micro-nano structured VNb9O25 anode with superior electronic conductivity for high-rate and long-life lithium storage. \emph{Journal of Materials Science \& Technology} \textbf{2021}, \emph{83}, 66--74\relax
\mciteBstWouldAddEndPuncttrue
\mciteSetBstMidEndSepPunct{\mcitedefaultmidpunct}
{\mcitedefaultendpunct}{\mcitedefaultseppunct}\relax
\EndOfBibitem
\bibitem[Ulutagay \latin{et~al.}(1998)Ulutagay, Schimek, and Hwu]{ulutagay1998niobium}
Ulutagay,~M.; Schimek,~G.~L.; Hwu,~S.-J. A niobium (V) arsenate: Nb9AsO25. \emph{Acta Crystallographica Section C: Crystal Structure Communications} \textbf{1998}, \emph{54}, 898--900\relax
\mciteBstWouldAddEndPuncttrue
\mciteSetBstMidEndSepPunct{\mcitedefaultmidpunct}
{\mcitedefaultendpunct}{\mcitedefaultseppunct}\relax
\EndOfBibitem
\bibitem[Fu \latin{et~al.}(2020)Fu, Cao, Liang, Luo, Chen, Murugadoss, Wu, Ding, Lin, and Guo]{fu2020highly}
Fu,~Q.; Cao,~H.; Liang,~G.; Luo,~L.; Chen,~Y.; Murugadoss,~V.; Wu,~S.; Ding,~T.; Lin,~C.; Guo,~Z. A highly Li+-conductive HfNb 24 O 62 anode material for superior Li+ storage. \emph{Chemical Communications} \textbf{2020}, \emph{56}, 619--622\relax
\mciteBstWouldAddEndPuncttrue
\mciteSetBstMidEndSepPunct{\mcitedefaultmidpunct}
{\mcitedefaultendpunct}{\mcitedefaultseppunct}\relax
\EndOfBibitem
\bibitem[Green \latin{et~al.}(2023)Green, Driscoll, Lakhdar, Kendrick, and Slater]{green2023structural}
Green,~A.; Driscoll,~E.; Lakhdar,~Y.; Kendrick,~E.; Slater,~P. Structural and electrochemical insights into novel Wadsley Roth Nb 7 Ti 1.5 Mo 1.5 O 25 and Ta 7 Ti 1.5 Mo 1.5 O 25 anodes for Li-ion battery application. \emph{Dalton Transactions} \textbf{2023}, \emph{52}, 13110--13118\relax
\mciteBstWouldAddEndPuncttrue
\mciteSetBstMidEndSepPunct{\mcitedefaultmidpunct}
{\mcitedefaultendpunct}{\mcitedefaultseppunct}\relax
\EndOfBibitem
\bibitem[Zhu \latin{et~al.}(2019)Zhu, Xu, Luo, Fu, Liang, Luo, Chen, Lin, and Zhao]{zhu2019monb}
Zhu,~X.; Xu,~J.; Luo,~Y.; Fu,~Q.; Liang,~G.; Luo,~L.; Chen,~Y.; Lin,~C.; Zhao,~X. MoNb 12 O 33 as a new anode material for high-capacity, safe, rapid and durable Li+ storage: structural characteristics, electrochemical properties and working mechanisms. \emph{Journal of Materials Chemistry A} \textbf{2019}, \emph{7}, 6522--6532\relax
\mciteBstWouldAddEndPuncttrue
\mciteSetBstMidEndSepPunct{\mcitedefaultmidpunct}
{\mcitedefaultendpunct}{\mcitedefaultseppunct}\relax
\EndOfBibitem
\bibitem[Patterson \latin{et~al.}(2023)Patterson, Elizalde-Segovia, Wyckoff, Zohar, Ding, Turner, Poeppelmeier, Narayan, Cl{\'e}ment, Seshadri, \latin{et~al.} others]{patterson2023rapid}
Patterson,~A.~R.; Elizalde-Segovia,~R.; Wyckoff,~K.~E.; Zohar,~A.; Ding,~P.~P.; Turner,~W.~M.; Poeppelmeier,~K.~R.; Narayan,~S.~R.; Cl{\'e}ment,~R.~J.; Seshadri,~R.; others Rapid and Reversible Lithium Insertion in the Wadsley--Roth-Derived Phase NaNb13O33. \emph{Chemistry of Materials} \textbf{2023}, \emph{35}, 6364--6373\relax
\mciteBstWouldAddEndPuncttrue
\mciteSetBstMidEndSepPunct{\mcitedefaultmidpunct}
{\mcitedefaultendpunct}{\mcitedefaultseppunct}\relax
\EndOfBibitem
\bibitem[Tao \latin{et~al.}(2022)Tao, Zhang, Tan, Jafta, Li, Liang, Sun, Wang, Fan, Lu, \latin{et~al.} others]{tao2022insight}
Tao,~R.; Zhang,~T.; Tan,~S.; Jafta,~C.~J.; Li,~C.; Liang,~J.; Sun,~X.-G.; Wang,~T.; Fan,~J.; Lu,~Z.; others Insight into the fast-rechargeability of a novel Mo1. 5W1. 5Nb14O44 anode material for high-performance lithium-ion batteries. \emph{Advanced Energy Materials} \textbf{2022}, \emph{12}, 2200519\relax
\mciteBstWouldAddEndPuncttrue
\mciteSetBstMidEndSepPunct{\mcitedefaultmidpunct}
{\mcitedefaultendpunct}{\mcitedefaultseppunct}\relax
\EndOfBibitem
\bibitem[Myslyvchenko \latin{et~al.}(2021)Myslyvchenko, Bondar, Tereshchenko, and Poliakov]{myslyvchenko2021formation}
Myslyvchenko,~O.; Bondar,~A.; Tereshchenko,~O.; Poliakov,~I. Formation of a new Wadsley-Roth phase during oxidation of Ti-Nb-Mo alloys. \emph{Materialia} \textbf{2021}, \emph{20}, 101213\relax
\mciteBstWouldAddEndPuncttrue
\mciteSetBstMidEndSepPunct{\mcitedefaultmidpunct}
{\mcitedefaultendpunct}{\mcitedefaultseppunct}\relax
\EndOfBibitem
\bibitem[Lin \latin{et~al.}(2015)Lin, Wang, Lin, Li, and Lu]{lin2015tinb}
Lin,~C.; Wang,~G.; Lin,~S.; Li,~J.; Lu,~L. TiNb 6 O 17: a new electrode material for lithium-ion batteries. \emph{Chemical Communications} \textbf{2015}, \emph{51}, 8970--8973\relax
\mciteBstWouldAddEndPuncttrue
\mciteSetBstMidEndSepPunct{\mcitedefaultmidpunct}
{\mcitedefaultendpunct}{\mcitedefaultseppunct}\relax
\EndOfBibitem
\bibitem[Wang \latin{et~al.}(1985)Wang, Greenblatt, and Kimura]{wang1985lithium}
Wang,~E.; Greenblatt,~M.; Kimura,~N. Lithium insertion in V/sub 3/Nb/sub 9/O/sub 29/. A Wadsley-Roth type phase. \emph{J. Electrochem. Soc.;(United States)} \textbf{1985}, \emph{132}\relax
\mciteBstWouldAddEndPuncttrue
\mciteSetBstMidEndSepPunct{\mcitedefaultmidpunct}
{\mcitedefaultendpunct}{\mcitedefaultseppunct}\relax
\EndOfBibitem
\bibitem[Wang \latin{et~al.}(2022)Wang, Yao, Li, Hu, Yin, Chen, Lu, Zhang, and Zhao]{wang2022fast}
Wang,~M.; Yao,~Z.; Li,~Q.; Hu,~Y.; Yin,~X.; Chen,~A.; Lu,~X.; Zhang,~J.; Zhao,~Y. Fast and extensive intercalation chemistry in Wadsley-Roth phase based high-capacity electrodes. \emph{Journal of Energy Chemistry} \textbf{2022}, \emph{69}, 601--611\relax
\mciteBstWouldAddEndPuncttrue
\mciteSetBstMidEndSepPunct{\mcitedefaultmidpunct}
{\mcitedefaultendpunct}{\mcitedefaultseppunct}\relax
\EndOfBibitem
\bibitem[Griffith \latin{et~al.}(2019)Griffith, Seymour, Hope, Butala, Lamontagne, Preefer, Ko{\c{c}}er, Henkelman, Morris, Cliffe, \latin{et~al.} others]{griffith2019ionic}
Griffith,~K.~J.; Seymour,~I.~D.; Hope,~M.~A.; Butala,~M.~M.; Lamontagne,~L.~K.; Preefer,~M.~B.; Ko{\c{c}}er,~C.~P.; Henkelman,~G.; Morris,~A.~J.; Cliffe,~M.~J.; others Ionic and electronic conduction in TiNb2O7. \emph{Journal of the American Chemical Society} \textbf{2019}, \emph{141}, 16706--16725\relax
\mciteBstWouldAddEndPuncttrue
\mciteSetBstMidEndSepPunct{\mcitedefaultmidpunct}
{\mcitedefaultendpunct}{\mcitedefaultseppunct}\relax
\EndOfBibitem
\bibitem[Preefer \latin{et~al.}(2020)Preefer, Saber, Wei, Bashian, Bocarsly, Zhang, Lee, Milam-Guerrero, Howard, Vincent, \latin{et~al.} others]{preefer2020multielectron}
Preefer,~M.~B.; Saber,~M.; Wei,~Q.; Bashian,~N.~H.; Bocarsly,~J.~D.; Zhang,~W.; Lee,~G.; Milam-Guerrero,~J.; Howard,~E.~S.; Vincent,~R.~C.; others Multielectron redox and insulator-to-metal transition upon lithium insertion in the fast-charging, Wadsley-Roth phase PNb9O25. \emph{Chemistry of Materials} \textbf{2020}, \emph{32}, 4553--4563\relax
\mciteBstWouldAddEndPuncttrue
\mciteSetBstMidEndSepPunct{\mcitedefaultmidpunct}
{\mcitedefaultendpunct}{\mcitedefaultseppunct}\relax
\EndOfBibitem
\bibitem[Saber \latin{et~al.}(2023)Saber, Behara, and Van~der Ven]{saber2023redox}
Saber,~M.; Behara,~S.~S.; Van~der Ven,~A. Redox Mechanisms, Structural Changes, and Electrochemistry of the Wadsley--Roth Li x TiNb2O7 Electrode Material. \emph{Chemistry of Materials} \textbf{2023}, \emph{35}, 9657--9668\relax
\mciteBstWouldAddEndPuncttrue
\mciteSetBstMidEndSepPunct{\mcitedefaultmidpunct}
{\mcitedefaultendpunct}{\mcitedefaultseppunct}\relax
\EndOfBibitem
\bibitem[Ko{\c{c}}er \latin{et~al.}(2019)Ko{\c{c}}er, Griffith, Grey, and Morris]{koccer2019first}
Ko{\c{c}}er,~C.~P.; Griffith,~K.~J.; Grey,~C.~P.; Morris,~A.~J. First-principles study of localized and delocalized electronic states in crystallographic shear phases of niobium oxide. \emph{Physical Review B} \textbf{2019}, \emph{99}, 075151\relax
\mciteBstWouldAddEndPuncttrue
\mciteSetBstMidEndSepPunct{\mcitedefaultmidpunct}
{\mcitedefaultendpunct}{\mcitedefaultseppunct}\relax
\EndOfBibitem
\bibitem[Saber \latin{et~al.}(2023)Saber, Reynolds, Li, Pollock, and Van~der Ven]{saber2023chemical}
Saber,~M.; Reynolds,~C.; Li,~J.; Pollock,~T.~M.; Van~der Ven,~A. Chemical and structural factors affecting the stability of Wadsley--Roth block phases. \emph{Inorganic Chemistry} \textbf{2023}, \emph{62}, 17317--17332\relax
\mciteBstWouldAddEndPuncttrue
\mciteSetBstMidEndSepPunct{\mcitedefaultmidpunct}
{\mcitedefaultendpunct}{\mcitedefaultseppunct}\relax
\EndOfBibitem
\bibitem[Jain \latin{et~al.}(2013)Jain, Ong, Hautier, Chen, Richards, Dacek, Cholia, Gunter, Skinner, Ceder, \latin{et~al.} others]{jain2013commentary}
Jain,~A.; Ong,~S.~P.; Hautier,~G.; Chen,~W.; Richards,~W.~D.; Dacek,~S.; Cholia,~S.; Gunter,~D.; Skinner,~D.; Ceder,~G.; others Commentary: The Materials Project: A materials genome approach to accelerating materials innovation. \emph{APL materials} \textbf{2013}, \emph{1}\relax
\mciteBstWouldAddEndPuncttrue
\mciteSetBstMidEndSepPunct{\mcitedefaultmidpunct}
{\mcitedefaultendpunct}{\mcitedefaultseppunct}\relax
\EndOfBibitem
\bibitem[McInnes \latin{et~al.}(2018)McInnes, Healy, and Melville]{mcinnes2018umap}
McInnes,~L.; Healy,~J.; Melville,~J. Umap: Uniform manifold approximation and projection for dimension reduction. \emph{arXiv preprint arXiv:1802.03426} \textbf{2018}, \relax
\mciteBstWouldAddEndPunctfalse
\mciteSetBstMidEndSepPunct{\mcitedefaultmidpunct}
{}{\mcitedefaultseppunct}\relax
\EndOfBibitem
\bibitem[Batatia \latin{et~al.}(2023)Batatia, Benner, Chiang, Elena, Kov{\'a}cs, Riebesell, Advincula, Asta, Baldwin, Bernstein, \latin{et~al.} others]{batatia2023foundation}
Batatia,~I.; Benner,~P.; Chiang,~Y.; Elena,~A.~M.; Kov{\'a}cs,~D.~P.; Riebesell,~J.; Advincula,~X.~R.; Asta,~M.; Baldwin,~W.~J.; Bernstein,~N.; others A foundation model for atomistic materials chemistry. \emph{arXiv preprint arXiv:2401.00096} \textbf{2023}, \relax
\mciteBstWouldAddEndPunctfalse
\mciteSetBstMidEndSepPunct{\mcitedefaultmidpunct}
{}{\mcitedefaultseppunct}\relax
\EndOfBibitem
\bibitem[Waroquiers \latin{et~al.}(2017)Waroquiers, Gonze, Rignanese, Welker-Nieuwoudt, Rosowski, Gobel, Schenk, Degelmann, Andr{\'e}, Glaum, \latin{et~al.} others]{waroquiers2017statistical}
Waroquiers,~D.; Gonze,~X.; Rignanese,~G.-M.; Welker-Nieuwoudt,~C.; Rosowski,~F.; Gobel,~M.; Schenk,~S.; Degelmann,~P.; Andr{\'e},~R.; Glaum,~R.; others Statistical analysis of coordination environments in oxides. \emph{Chemistry of Materials} \textbf{2017}, \emph{29}, 8346--8360\relax
\mciteBstWouldAddEndPuncttrue
\mciteSetBstMidEndSepPunct{\mcitedefaultmidpunct}
{\mcitedefaultendpunct}{\mcitedefaultseppunct}\relax
\EndOfBibitem
\bibitem[Sturgill \latin{et~al.}(2025)Sturgill, Sutton, Schwenzel, and Stefik]{sturgill2025capacity}
Sturgill,~C.; Sutton,~C.; Schwenzel,~J.; Stefik,~M. Capacity-Weighted Figures-of-Merit for Battery Transport Metrics. \emph{Journal of Materials Chemistry A} \textbf{2025}, \relax
\mciteBstWouldAddEndPunctfalse
\mciteSetBstMidEndSepPunct{\mcitedefaultmidpunct}
{}{\mcitedefaultseppunct}\relax
\EndOfBibitem
\bibitem[Kresse and Hafner(1993)Kresse, and Hafner]{kresse1993ab}
Kresse,~G.; Hafner,~J. Ab initio molecular dynamics for liquid metals. \emph{Physical review B} \textbf{1993}, \emph{47}, 558\relax
\mciteBstWouldAddEndPuncttrue
\mciteSetBstMidEndSepPunct{\mcitedefaultmidpunct}
{\mcitedefaultendpunct}{\mcitedefaultseppunct}\relax
\EndOfBibitem
\bibitem[Kresse and Furthm{\"u}ller(1996)Kresse, and Furthm{\"u}ller]{kresse1996efficiency}
Kresse,~G.; Furthm{\"u}ller,~J. Efficiency of ab-initio total energy calculations for metals and semiconductors using a plane-wave basis set. \emph{Computational materials science} \textbf{1996}, \emph{6}, 15--50\relax
\mciteBstWouldAddEndPuncttrue
\mciteSetBstMidEndSepPunct{\mcitedefaultmidpunct}
{\mcitedefaultendpunct}{\mcitedefaultseppunct}\relax
\EndOfBibitem
\bibitem[Kresse and Furthm{\"u}ller(1996)Kresse, and Furthm{\"u}ller]{kresse1996efficient}
Kresse,~G.; Furthm{\"u}ller,~J. Efficient iterative schemes for ab initio total-energy calculations using a plane-wave basis set. \emph{Physical review B} \textbf{1996}, \emph{54}, 11169\relax
\mciteBstWouldAddEndPuncttrue
\mciteSetBstMidEndSepPunct{\mcitedefaultmidpunct}
{\mcitedefaultendpunct}{\mcitedefaultseppunct}\relax
\EndOfBibitem
\bibitem[Perdew \latin{et~al.}(1996)Perdew, Burke, and Ernzerhof]{perdew1996generalized}
Perdew,~J.~P.; Burke,~K.; Ernzerhof,~M. Generalized gradient approximation made simple. \emph{Physical review letters} \textbf{1996}, \emph{77}, 3865\relax
\mciteBstWouldAddEndPuncttrue
\mciteSetBstMidEndSepPunct{\mcitedefaultmidpunct}
{\mcitedefaultendpunct}{\mcitedefaultseppunct}\relax
\EndOfBibitem
\bibitem[Kresse and Hafner(1994)Kresse, and Hafner]{kresse1994norm}
Kresse,~G.; Hafner,~J. Norm-conserving and ultrasoft pseudopotentials for first-row and transition elements. \emph{Journal of Physics: Condensed Matter} \textbf{1994}, \emph{6}, 8245\relax
\mciteBstWouldAddEndPuncttrue
\mciteSetBstMidEndSepPunct{\mcitedefaultmidpunct}
{\mcitedefaultendpunct}{\mcitedefaultseppunct}\relax
\EndOfBibitem
\bibitem[Kresse and Joubert(1999)Kresse, and Joubert]{kresse1999ultrasoft}
Kresse,~G.; Joubert,~D. From ultrasoft pseudopotentials to the projector augmented-wave method. \emph{Physical review b} \textbf{1999}, \emph{59}, 1758\relax
\mciteBstWouldAddEndPuncttrue
\mciteSetBstMidEndSepPunct{\mcitedefaultmidpunct}
{\mcitedefaultendpunct}{\mcitedefaultseppunct}\relax
\EndOfBibitem
\bibitem[Ong \latin{et~al.}(2015)Ong, Cholia, Jain, Brafman, Gunter, Ceder, and Persson]{ong2015materials}
Ong,~S.~P.; Cholia,~S.; Jain,~A.; Brafman,~M.; Gunter,~D.; Ceder,~G.; Persson,~K.~A. The Materials Application Programming Interface (API): A simple, flexible and efficient API for materials data based on REpresentational State Transfer (REST) principles. \emph{Computational Materials Science} \textbf{2015}, \emph{97}, 209--215\relax
\mciteBstWouldAddEndPuncttrue
\mciteSetBstMidEndSepPunct{\mcitedefaultmidpunct}
{\mcitedefaultendpunct}{\mcitedefaultseppunct}\relax
\EndOfBibitem
\bibitem[Ong \latin{et~al.}(2013)Ong, Richards, Jain, Hautier, Kocher, Cholia, Gunter, Chevrier, Persson, and Ceder]{ong2013python}
Ong,~S.~P.; Richards,~W.~D.; Jain,~A.; Hautier,~G.; Kocher,~M.; Cholia,~S.; Gunter,~D.; Chevrier,~V.~L.; Persson,~K.~A.; Ceder,~G. Python Materials Genomics (pymatgen): A robust, open-source python library for materials analysis. \emph{Computational Materials Science} \textbf{2013}, \emph{68}, 314--319\relax
\mciteBstWouldAddEndPuncttrue
\mciteSetBstMidEndSepPunct{\mcitedefaultmidpunct}
{\mcitedefaultendpunct}{\mcitedefaultseppunct}\relax
\EndOfBibitem
\bibitem[Wang \latin{et~al.}(2006)Wang, Maxisch, and Ceder]{wang2006oxidation}
Wang,~L.; Maxisch,~T.; Ceder,~G. Oxidation energies of transition metal oxides within the GGA+ U framework. \emph{Physical Review B} \textbf{2006}, \emph{73}, 195107\relax
\mciteBstWouldAddEndPuncttrue
\mciteSetBstMidEndSepPunct{\mcitedefaultmidpunct}
{\mcitedefaultendpunct}{\mcitedefaultseppunct}\relax
\EndOfBibitem
\bibitem[Wang \latin{et~al.}(2021)Wang, Kingsbury, McDermott, Horton, Jain, Ong, Dwaraknath, and Persson]{wang2021framework}
Wang,~A.; Kingsbury,~R.; McDermott,~M.; Horton,~M.; Jain,~A.; Ong,~S.~P.; Dwaraknath,~S.; Persson,~K.~A. A framework for quantifying uncertainty in DFT energy corrections. \emph{Scientific reports} \textbf{2021}, \emph{11}, 15496\relax
\mciteBstWouldAddEndPuncttrue
\mciteSetBstMidEndSepPunct{\mcitedefaultmidpunct}
{\mcitedefaultendpunct}{\mcitedefaultseppunct}\relax
\EndOfBibitem
\bibitem[Roth and Wadsley(1965)Roth, and Wadsley]{roth1965mixed}
Roth,~R.; Wadsley,~A. Mixed oxides of titanium and niobium: the crystal structure of TiNb24O62 (TiO2. 12Nb2O5). \emph{Acta Crystallographica} \textbf{1965}, \emph{18}, 724--730\relax
\mciteBstWouldAddEndPuncttrue
\mciteSetBstMidEndSepPunct{\mcitedefaultmidpunct}
{\mcitedefaultendpunct}{\mcitedefaultseppunct}\relax
\EndOfBibitem
\bibitem[Wadsley(1961)]{wadsley1961mixed}
Wadsley,~A. Mixed oxides of titanium and niobium. II. The crystal structures of the dimorphic forms Ti2Nb10O29. \emph{Acta Crystallographica} \textbf{1961}, \emph{14}, 664--670\relax
\mciteBstWouldAddEndPuncttrue
\mciteSetBstMidEndSepPunct{\mcitedefaultmidpunct}
{\mcitedefaultendpunct}{\mcitedefaultseppunct}\relax
\EndOfBibitem
\bibitem[Roth and Wadsley(1965)Roth, and Wadsley]{roth1965multiple}
Roth,~R.; Wadsley,~A. Multiple phase formation in the binary system Nb2O5--WO3. I. Preparation and identification of phases. \emph{Acta Crystallographica} \textbf{1965}, \emph{19}, 26--32\relax
\mciteBstWouldAddEndPuncttrue
\mciteSetBstMidEndSepPunct{\mcitedefaultmidpunct}
{\mcitedefaultendpunct}{\mcitedefaultseppunct}\relax
\EndOfBibitem
\bibitem[Voskanyan \latin{et~al.}(2020)Voskanyan, Abramchuk, and Navrotsky]{voskanyan2020entropy}
Voskanyan,~A.~A.; Abramchuk,~M.; Navrotsky,~A. Entropy stabilization of TiO2--Nb2O5 Wadsley--Roth shear phases and their prospects for lithium-ion battery anode materials. \emph{Chemistry of Materials} \textbf{2020}, \emph{32}, 5301--5308\relax
\mciteBstWouldAddEndPuncttrue
\mciteSetBstMidEndSepPunct{\mcitedefaultmidpunct}
{\mcitedefaultendpunct}{\mcitedefaultseppunct}\relax
\EndOfBibitem
\bibitem[Zunger \latin{et~al.}(1990)Zunger, Wei, Ferreira, and Bernard]{zunger1990special}
Zunger,~A.; Wei,~S.-H.; Ferreira,~L.~G.; Bernard,~J.~E. Special quasirandom structures. \emph{Physical review letters} \textbf{1990}, \emph{65}, 353\relax
\mciteBstWouldAddEndPuncttrue
\mciteSetBstMidEndSepPunct{\mcitedefaultmidpunct}
{\mcitedefaultendpunct}{\mcitedefaultseppunct}\relax
\EndOfBibitem
\bibitem[van~de Walle \latin{et~al.}(2013)van~de Walle, Tiwary, de~Jong, Olmsted, Asta, Dick, Shin, Wang, Chen, and Liu]{van2013efficient}
van~de Walle,~A.; Tiwary,~P.; de~Jong,~M.; Olmsted,~D.~L.; Asta,~M.; Dick,~A.; Shin,~D.; Wang,~Y.; Chen,~L.-Q.; Liu,~Z.-K. Efficient stochastic generation of special quasirandom structures. \emph{Calphad} \textbf{2013}, \emph{42}, 13--18\relax
\mciteBstWouldAddEndPuncttrue
\mciteSetBstMidEndSepPunct{\mcitedefaultmidpunct}
{\mcitedefaultendpunct}{\mcitedefaultseppunct}\relax
\EndOfBibitem
\bibitem[Hautier \latin{et~al.}()Hautier, Fischer, Ehrlacher, Jain, and Ceder]{hautier_data_2011}
Hautier,~G.; Fischer,~C.; Ehrlacher,~V.; Jain,~A.; Ceder,~G. Data Mined Ionic Substitutions for the Discovery of New Compounds. \emph{50}, 656--663, Publisher: American Chemical Society\relax
\mciteBstWouldAddEndPuncttrue
\mciteSetBstMidEndSepPunct{\mcitedefaultmidpunct}
{\mcitedefaultendpunct}{\mcitedefaultseppunct}\relax
\EndOfBibitem
\bibitem[Chien \latin{et~al.}(2023)Chien, Liu, Menon, Brant, Brandell, and Lacey]{chien2023rapid}
Chien,~Y.-C.; Liu,~H.; Menon,~A.~S.; Brant,~W.~R.; Brandell,~D.; Lacey,~M.~J. Rapid determination of solid-state diffusion coefficients in Li-based batteries via intermittent current interruption method. \emph{Nature communications} \textbf{2023}, \emph{14}, 2289\relax
\mciteBstWouldAddEndPuncttrue
\mciteSetBstMidEndSepPunct{\mcitedefaultmidpunct}
{\mcitedefaultendpunct}{\mcitedefaultseppunct}\relax
\EndOfBibitem
\end{mcitethebibliography}

\end{document}